\documentclass[10pt]{asme2ej}

\usepackage{setspace}
\usepackage{graphicx} 
\usepackage{amsmath,amssymb}
\usepackage{hyperref}   
\hypersetup{
	colorlinks=true,
	linkcolor=blue,
	citecolor=blue,
	urlcolor=blue,
}
\usepackage{xcolor}
\usepackage[normalem]{ulem}
\usepackage[square,numbers]{natbib}

\title{Improvement of Reduced-Order Model \textcolor{black}{for Two-Dimensional Cylinder Flow} Based on Global Proper Orthogonal Decomposition in terms of Robustness and Computational Speed}

\author{Yuto Nakamura\thanks{Address all correspondence related to ASME style format and figures to this author.}
    \affiliation{
	Ph.D. Student\\
	Propulsion Engineering Laboratory\\
	Department of Aerospace Engineering\\
	Tohoku University\\
	Sendai, 980-8579, Miyagi, Japan\\
    yuto.nakamura.t4@dc.tohoku.ac.jp
    }	
}

\author{Shintaro Sato
    \affiliation{
	Assistant Professor \\
	Propulsion Engineering Laboratory\\
	Department of Aerospace Engineering\\
	Tohoku University\\
	Sendai, 980-8579, Miyagi, Japan
    }	
}
\author{Naofumi Ohnishi
    \affiliation{
	Professor \\
	Propulsion Engineering Laboratory\\
	Department of Aerospace Engineering\\
	Tohoku University\\
	Sendai, 980-8579, Miyagi, Japan
    }	
}

\begin{document}

\maketitle    


\begin{abstract}
Reduced-order models (ROMs) are widely used in fluid engineering to enable rapid prediction of flow fields for parametric analysis, design optimization, and control applications. Proper orthogonal decomposition (POD) is commonly employed to construct ROMs because it provides an optimal basis for representing a given flow dataset. However, POD-based ROMs often lack robustness when applied to flow conditions that differ from those included in the training data. Incorporating multiple flow conditions can improve robustness, but this generally increases the computational cost of ROM prediction, which limits practical applicability in engineering workflows.
In this study, we propose a ROM framework that achieves fast and robust flow prediction even when the dataset contains a large number of flow conditions. The proposed approach employs a novel two-step order-reduction strategy based on POD. In the second reduction step, flow conditions that are most relevant to the target prediction are selectively retained, thereby reducing the computational cost without sacrificing accuracy. The performance of the proposed ROM is evaluated for a two-dimensional unsteady flow past a circular cylinder, a canonical benchmark problem in fluid engineering. The model accurately reproduces the relationship between vortex-shedding frequency and Reynolds number obtained from full numerical simulations. Furthermore, the proposed ROM reduces the computational cost by approximately 50\% compared with a conventional POD-based ROM constructed using flow data at $27$ different Reynolds numbers.
\end{abstract}



\section{Introduction}
Computational fluid dynamics (CFD) has become an indispensable tool in both fluid mechanics research and engineering applications\cite{CFD1,CFD2,aeroreview,Iwatani,Deniz_2022}. Owing to the rapid improvement of computer performance and numerical algorithms, CFD has been widely applied to flow control\cite{control1,control2,Liu_2020} and aerodynamic design\cite{design1,design2,design3}. In these applications, engineers often need to evaluate how the flow field depends on various parameters such as the Reynolds number\cite{Re1,Re2} or control inputs for flow-control devices like plasma actuators\cite{PA1,PA3,Pendar_2022, flowcontrol1,flowcontrol2,flowcontrol3,flowcontrol4,Murugan_2021}.

Despite its accuracy, CFD demands extensive computational resources because multiple simulations are required to explore different flow conditions\cite{times1,times2,times3}. Obtaining a complete dataset for design and control purposes thus incurs enormous computational cost. Furthermore, flow-control applications require rapid predictions of flow variations in response to control inputs, making conventional CFD too time-consuming for practical use. Consequently, there is a growing demand for fast numerical models capable of predicting flow behavior efficiently while maintaining reasonable accuracy.

A reduced-order model (ROM)\cite{ROM1,ROM2,ROM3} offers an attractive alternative by enabling fast computation of approximate flow fields. Unlike full-order numerical simulations, ROMs focus on capturing the dominant dynamics rather than reproducing the entire flow field with high fidelity. Similar to surrogate models, ROMs are empirically constructed from flow datasets obtained through high-fidelity simulations. However, a persistent challenge is that such data-driven ROMs often struggle to predict flows under conditions not included in the training dataset, as their formulation depends strongly on the empirical data.

Proper orthogonal decomposition (POD)\cite{POD0,Holmes,POD1,POD15,POD16,POD17,POD2,POD_snap,Kawaguchi,Munir_2022,Abdul_Salam_2024,Hayden_2024,Kambayashi_2025} has been widely used to extract dominant spatial structures from time-dependent flow fields and serves as a foundation for many ROMs. The POD provides an optimal set of orthogonal modes that best represent the given flow data. 
In this way, the POD basis provides a low-dimensional representation of a high-dimensional flow field. This representation can be used both to estimate the entire flow field\cite{Hayden_2024, Kambayashi_2025} and to identify the dominant flow dynamics\cite{Munir_2022,Abdul_Salam_2024}. 
By projecting the dynamics onto a low-dimensional subspace spanned by a small number of POD modes, the resulting ROM achieves substantial computational efficiency compared with conventional CFD\cite{POD_Galerkin,shift,NACAGP,GP_deane}. 
Such low-dimensionalization constitutes one of the most fundamental approaches in ROM, and many ROMs in fluid mechanics are constructed based on the concept of projection.
Nevertheless, the POD basis obtained from one specific condition is not guaranteed to be optimal for flow fields at different conditions. When the Reynolds number or other parameters change, the low-dimensional space derived from the original dataset may fail to reconstruct the new flow field, as essential modes may be missing.

To address this limitation, several studies have extended the POD-based ROM to datasets containing multiple flow conditions\cite{POD_ANN,POD_RBF,cylinder7}. Performing POD on such a multi-condition dataset, often termed global POD, enables the construction of a more robust ROM that can predict flows beyond the original dataset\cite{GLOBALPOD0,GLOBALPOD1,GLOBALPOD2}. The global POD-based ROM retains the simplicity of the conventional POD approach while offering improved generalization across conditions. However, as the number of included conditions increases, the diversity of flow features also grows, requiring a larger number of POD modes to achieve sufficient accuracy. This expansion in modal dimension increases the computational cost, thereby offsetting the efficiency advantage of ROMs.

In this study, we propose a modified ROM framework that maintains robustness to varying flow conditions while preserving computational efficiency.
\textcolor{black}{As a strategy to construct robust ROMs based on global POD, a method has been proposed in which the dataset is divided into multiple subsets, and POD is performed separately for each subset\cite{chaudhry2017exploration}. In the prediction stage, the POD basis corresponding to the subset closest to the target condition is selected, enabling condition-adaptive basis selection. Other studies have also reported that the choice of conditions included in the dataset is closely related to the predictive accuracy of the ROM\cite{karcher2022adaptive}, suggesting that selecting appropriate training conditions from the dataset contributes to the improvement of robustness. In addition, several approaches have been proposed to reduce the computational cost of POD basis construction or to decrease memory requirements during the online phase by performing POD in multiple stages, either hierarchically or incrementally\cite{himpe2018hierarchical,Ross_2008}. These approaches contribute both to improving the robustness of POD-based ROMs and to reducing computational cost.}

 The proposed method divides the global POD into two parts to achieve faster prediction without sacrificing accuracy. Importantly, the algorithm remains simple to implement, involving no additional complex operations beyond the conventional POD procedure. The numerical methods used to generate the datasets are described in Section~\ref{numerical}. The conventional global POD-based ROM is formulated and demonstrated for two-dimensional quasi-steady flow around a cylinder in Sections~\ref{globalmethod} and~\ref{globalapplication}. Based on these results, the proposed ROM is formulated in Section~\ref{dualmethod} and validated in Section~\ref{dualapplication}, followed by concluding remarks in Section~\ref{conclusion}.

\section{Numerical Method for POD Dataset}\label{numerical}
The flow fields around a circular cylinder are computed by numerical simulation for the dataset. The governing equations are the two-dimensional incompressible Navier--Stokes and continuity equations presented below:
\begin{equation}
\nabla\cdot\boldsymbol{u}=0,
   \label{eqcont}	
\end{equation}
\begin{equation}
\frac{\partial{\boldsymbol{u}}}{\partial{t}}=-\nabla\cdot\boldsymbol{u}\boldsymbol{u}-\frac{1}{\rho}\nabla{p}+{\nu}{\nabla^2}\boldsymbol{u},
   \label{eqnavi}
\end{equation}
where $\boldsymbol{u}$ represents the vector of the velocity (bold symbol represents vectors), ${p}$ is the pressure, $\nu$ is the kinematic viscosity, and $\rho$ is the fluid density. The Reynolds number is defined using the mainstream velocity $U_{\infty}$ and characteristic length $D$ as follows
\begin{equation}
Re=\frac{U_{\infty}D}{\nu}.
   \label{eqRe}
\end{equation}
The governing equations are non-dimensionalized and transformed into a curvilinear coordinate system
 \begin{equation}
\sum^2_{j=1}\frac{1}{J}\frac{\partial}{\partial{\xi^{j}}}\left({JU}^{j}\right)=0,
   \label{eqUcont}
\end{equation}
\begin{equation}
\begin{split}
\frac{\partial{{u}_i}}{\partial{t}}&=
-\sum^2_{j=1}\frac{1}{J}\frac{\partial}{\partial{\xi^{j}}}\left({JU}^{j}{{u}_i}\right)\\
&+\sum^2_{j=1}\frac{1}{\rho}\frac{\partial\xi^{j}}{\partial{x_{i}}}\frac{\partial{p}}{\partial{\xi^{j}}}\\
&+\sum^2_{j=1}\frac{1}{Re}\frac{1}{J}\frac{\partial}{\partial{\xi^{j}}}\sum^2_{k=1}\left({\gamma^{jk}\frac{\partial{{u}_i}}{\partial{\xi^{k}}}}\right)
\,\,\,(i=1,2),
   \label{eqUnavi}
\end{split}
\end{equation}
where 
\begin{equation}
\gamma^{jk}=\sum^2_{l=1}J\frac{\partial{\xi^{j}}}{\partial{x_{l}}}\frac{\partial{\xi^{k}}}{\partial{x_{l}}},
   \label{eqtensor}
\end{equation}
represents a symmetric tensor, ${u_i}$ is the $i$th component of Cartesian velocity vector, ${U^j}$ is the $j$th component of contravariant component vector. 
Here, the advection, viscosity, and pressure terms are respectively defined by the operators
\begin{equation}
\begin{split}
A(J\boldsymbol{U},{u}_i)=\sum^2_{j=1}\frac{1}{J}\frac{\partial}{\partial{\xi^{j}}}\left({JU}^{j}{{u}_i}\right),
   \label{eqadvec}
\end{split}
\end{equation}
\begin{equation}
\begin{split}
B({u}_i)=\sum^2_{j=1}\frac{1}{J}\frac{\partial}{\partial{\xi^{j}}}\sum^2_{k=1}\left({\gamma^{jk}\frac{\partial{{u}_i}}{\partial{\xi^{k}}}}\right),
   \label{eqvis}
\end{split}
\end{equation}
\begin{equation}
\begin{split}
P_i(p)=\sum^2_{j=1}\frac{\partial\xi^{j}}{\partial{x_{i}}}\frac{\partial{p}}{\partial{\xi^{j}}},
   \label{eqpre}
\end{split}
\end{equation}
where $J\boldsymbol{U}$ represents
\begin{equation}
\begin{split}
J\boldsymbol{U}=\begin{bmatrix}
    \displaystyle{JU^1}\\
    \displaystyle{JU^2}\\
\end{bmatrix}.
   \label{eqpre}
\end{split}
\end{equation}
The \textcolor{black}{Navier}-Stokes equation is rewritten as
\begin{equation}
\begin{split}
\frac{\partial{{u}_i}}{\partial{t}}=
-A(J\boldsymbol{U},{u}_i)
+\frac{1}{\rho}P_i(p)
+\frac{1}{Re}B({u}_i).
   \label{eqUnavi_byoperator}
\end{split}
\end{equation}
Equations (\ref{eqUcont}) and (\ref{eqUnavi}) are solved by an algorithm that extends the fractional step method proposed by Le and Moin\cite{RKincomp} to generalized coordinates. The governing equations are solved in the following three steps
\begin{enumerate}
   \item[(a)] The Navier--Stokes equation, without the pressure term, is computed to predict velocity ${\boldsymbol{u}}'$ and contravariant velocity $J\boldsymbol{U}'$.
   \item[(b)] Pressure potential $\phi$ is computed using the predicted contravariant velocity $J\boldsymbol{U}'$.
   \item[(c)] Using the pressure potential $\phi$, the velocity ${u}'_i$ and contravariant velocity $J\boldsymbol{U}'$ are corrected to satisfy the continuity equation.
\end{enumerate}

The Le and Moin's algorithm uses the third-order third-stage Runge--Kutta scheme for the advection term and the second-order implicit Crank--Nicolson scheme for the viscous term in the time advancement of step (a). The three steps in Runge--Kutta time advancement are written as follows
\begin{equation}
\begin{split}
&\frac{{\hat{u}_i^k}-{\hat{u}_i^{k-1}}}{\Delta{t}}
=-\zeta_kA(J\hat{\boldsymbol{U}}^{*k-1},\hat{u}_i^{*k-1})\\
&\,\,\,\,\,\,\,\,\,\,\,\,\,\,\,\,\,\,\,\,\,\,\,\,\,\,\,\,\,\,\,\,\,\,\,\,\,\,\,\,\,\,\,
-\eta_kA(J\hat{\boldsymbol{U}}^{*k-2},\hat{u}_i^{*k-2})\\
+&\frac{({\alpha_k+\beta_k})}{Re}B(\hat{u}_i^{k-1})+\frac{\beta_k}{Re}B(\hat{u}_i^{k}-\hat{u}_i^{k-1})\\
&\,\,\,\,\,\,\,\,\,\,\,\,\,\,\,\,\,\,\,\,\,\,\,\,\,\,\,\,\,\,\,\,\,\,\,\,\,\,\,\,
-\frac{\alpha_k}{Re}C_i(\hat{\boldsymbol{u}}^{k-1})
\,\,\,(k=1,2,3),
   \label{eqRKstep}
\end{split}
\end{equation}
\begin{equation}
\begin{split}
\hat{u}_i^{*k}=\hat{u}_i^k-{\Delta{t}}\sum^k_{l=1}(\alpha_l+\beta_l)P_i(\phi^{n})\,\,\,(k=1,2),
   \label{eqRKstep}
\end{split}
\end{equation}
\begin{equation}
\begin{split}
{u}'_i=\hat{u}^3_i,
\label{eqadvec}
\end{split}
\end{equation}
\begin{equation}
\begin{split}
J\hat{\boldsymbol{U}}^{*k}=\boldsymbol{D}(\hat{\boldsymbol{u}}^{k})-{\Delta{t}}\sum^k_{l=1}(\alpha_l+\beta_l)&\boldsymbol{Q}(\phi^{n})\\
&(k=1,2),
\label{eqadvec}
\end{split}
\end{equation}
\begin{equation}
\begin{split}
J{\boldsymbol{U}}^{'}=\boldsymbol{D}(\hat{\boldsymbol{u}}^{3}),
\label{eqadvec}
\end{split}
\end{equation}
where $\hat{u}_i^{k}$ represents the vector component of the kth step in the Runge--Kutta method, $\phi^{n}$ is the solution of the pressure equation in step (c) for the previous time step $n$, 
\begin{equation}
\begin{split}
\hat{u}^0_i={u}^n_i,\,\,\,\hat{u}^{*0}_i={u}^n_i,\,\,\,J\hat{\boldsymbol{U}}^{*0}=J{\boldsymbol{U}}^{n},
\label{eqadvec}
\end{split}
\end{equation}
coefficients are
\begin{equation}
\begin{split}
\zeta_1=\frac{8}{15},\,\,\,&\zeta_2=\frac{5}{12},\,\,\,\zeta_{\textcolor{black}{3}}=\frac{3}{4},\\
\eta_1=0,\,\,\,&\eta_2=-\frac{17}{60},\,\,\,\eta_3=\frac{5}{12},\\
&\alpha_1=\beta_1=\frac{4}{15},\\
&\alpha_2=\beta_2=\frac{1}{15},\\
&\alpha_3=\beta_3=\frac{1}{6},\\
\label{eqadvec}
\end{split}
\end{equation}
and operators $C_i(\boldsymbol{u})$, $\boldsymbol{D}(\boldsymbol{u})$, and $\boldsymbol{Q}(\phi)$ is presented below
\begin{equation}
\begin{split}
&C_i(\boldsymbol{u})\\
&\,=\sum^2_{l=1}\frac{1}{J}\frac{\partial{\xi^{l}}}{\partial{x_{i}}}\frac{\partial}{\partial{\xi^{l}}}
\left\{\sum^2_{m=1}\frac{\partial}{\partial{\xi^{m}}}\left(\sum^2_{j=1}J\frac{\partial{\xi^{m}}}{\partial{x_{j}}}{u_j}\right)\right\}\,\,\,\\
&\,\,\,\,\,\,\,\,\,\,\,\,\,\,\,\,\,\,\,\,\,\,\,\,\,\,\,\,\,\,\,\,\,\,\,\,\,\,\,\,\,\,\,\,\,\,\,\,\,\,\,\,\,\,\,\,\,\,\,\,\,\,\,\,\,\,\,\,\,\,\,\,\,\,\,\,\,\,\,\,\,\,\,\,\,\,\,\,\,\,\,\,\,\,\,\,(i=1,2),
   \label{eqC}
\end{split}
\end{equation}
\begin{equation}
\begin{split}
\boldsymbol{D}(\boldsymbol{u})=
\begin{bmatrix}
    \displaystyle{{D_1}(\boldsymbol{u})}\\
    \displaystyle{{D_2}(\boldsymbol{u})}\\
\end{bmatrix}=
\begin{bmatrix}
    \displaystyle{\sum^2_{j=1}J\frac{\partial{\xi^{1}}}{\partial{x_{j}}}{u_j}}\\
    \displaystyle{\sum^2_{j=1}J\frac{\partial{\xi^{2}}}{\partial{x_{j}}}{u_j}}\\
\end{bmatrix},
   \label{eqD}
\end{split}
\end{equation}
\begin{equation}
\begin{split}
\boldsymbol{Q}(\phi)=
\begin{bmatrix}
    \displaystyle{{Q_1}(\phi)}\\
    \displaystyle{{Q_2}(\phi)}\\
\end{bmatrix}=
\begin{bmatrix}
    \displaystyle{\gamma^{1j}\frac{\partial{\phi}}{\partial{\xi^{j}}}}\\
    \displaystyle{\gamma^{2j}\frac{\partial{\phi}}{\partial{\xi^{j}}}}\\
\end{bmatrix},
   \label{eqQ}
\end{split}
\end{equation}
respectively. The advection term $A$ is discretized by the QUICK method\cite{QUICK}, and the other spatial differences are evaluated by second-order central difference. The theoretical maximum value of the CFL number in this time integration method is $\sqrt{3}$. In this study, the CFL number remains below or equal to $0.8$. 

In the step (b), $\phi^{n+1}$ is obtained by solving the following equation
\begin{equation}
\begin{split}
JB(\phi^{n+1})=\frac{1}{\Delta t}\sum^2_{j=1}\frac{\partial}{\partial{x_{j}}}J{\boldsymbol{U}}^{'}.
\label{eqadvec}
\end{split}
\end{equation}
This equation is solved by the bi-conjugate gradient stabilized (BiCGSTAB) method\cite{BiCGstab}. An incomplete LU\cite{iLU1,iLU2} decomposition is used as a preconditioner. 

In the step (c), the predicted velocity $\boldsymbol{u}'$ and contravariant velocity $J\boldsymbol{U}'$ are corrected by the potential $\phi^{n+1}$ as follows
\begin{equation}
\begin{split}
{u}_i^{n+1}=\hat{u}'_i-{\Delta{t}}P_i(\phi^{n+1}),
   \label{equcorect}
\end{split}
\end{equation}
\begin{equation}
\begin{split}
J{\boldsymbol{U}}^{n+1}=J\hat{\boldsymbol{U}}'-{\Delta{t}}\boldsymbol{Q}(\phi^{n+1}).
\label{eqJUcorect}
\end{split}
\end{equation}
The pressure is computed from the pressure potential presented below
\begin{equation}
p^{n+1}={\rho}\phi^{n+1}-{\rho}\frac{\Delta{t}}{2}B(\phi^{n+1}).
   \label{eqphosei}
\end{equation}

\begin{figure}[!ht]
   \centering
   \includegraphics[scale=0.4]{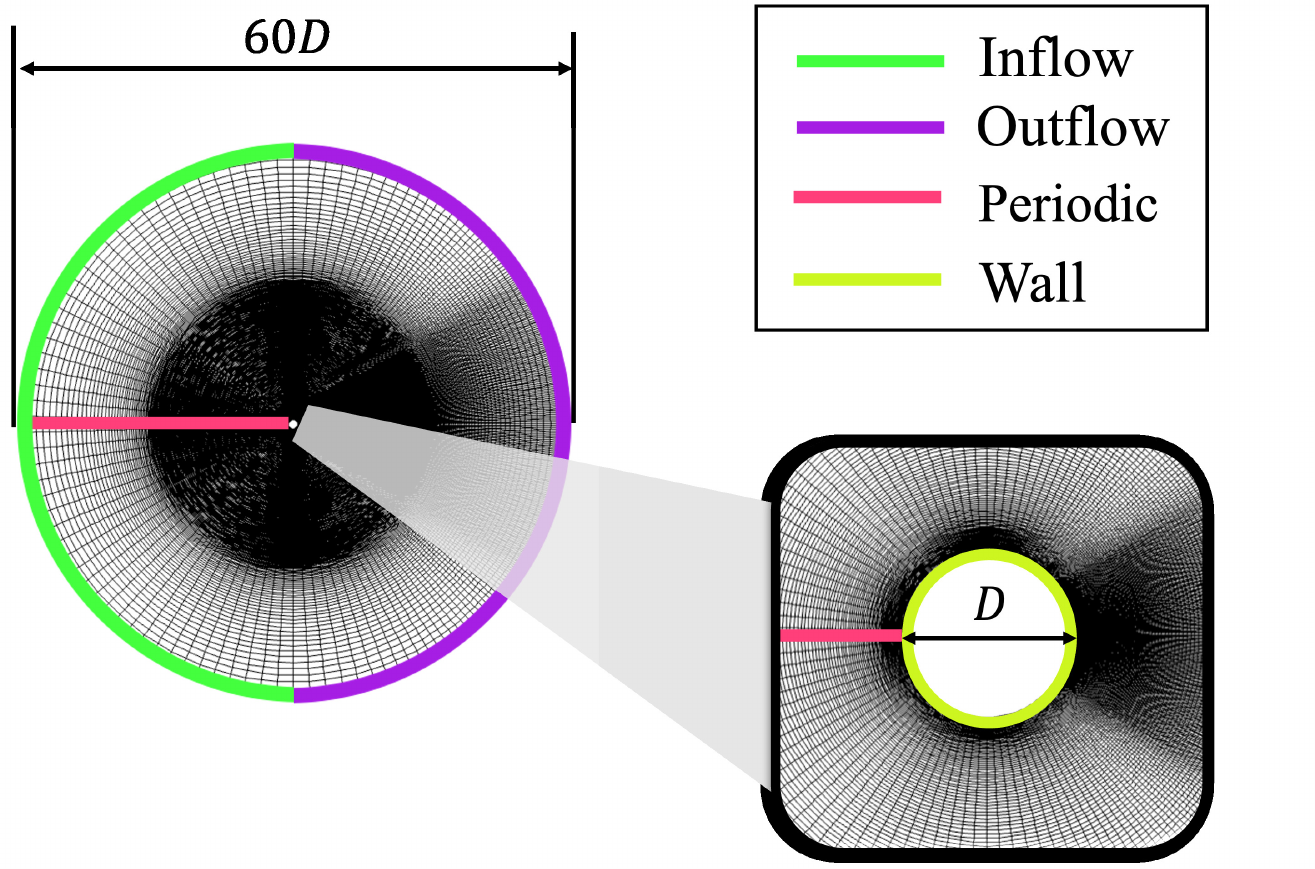}
   \caption{Computational grid and boundary conditions for flow around a circular cylinder. }
   \label{gridandboundary}
\end{figure}

The computational grid around a circular cylinder is an O-type grid. The far-field boundary of the computational grid extends up to 60 times the diameter of a circular cylinder, as shown in Fig. \ref{gridandboundary}. 
The number of grids is $280$ in the wall-normal direction and $480$ in the wall-parallel direction. The height of the first layer next to the cylinder is $\Delta{x} = 10^{-3}D$. The maximum cell expansion ratio in all domains is 1.1. These grid parameters are based on Jiang et al.\cite{cylinder_DNS}. The Dirichlet condition gives the boundary conditions for the velocity at the inflow and wall conditions. The \textcolor{black}{Neumann} condition gives the boundary conditions for the pressure at the inflow and wall conditions. The boundary condition proposed by Gresho\cite{outbound} is used for the outflow boundary.

\section{Global POD-Based ROM}\label{globalmethod}
\subsection{POD Formulation}\label{cPOD}
POD is a method for computing the orthogonal modes that best represent a given dataset. Let vector-valued dataset $\boldsymbol{q}_m\,\,\,(m=1,2,3,\cdots,m_{\text{max}})$ dependent on a set of independent variable $\boldsymbol{x}$ be given. All $m_{\text{max}}$ data belong to a Hilbert space $\mathcal{H}$ whose inner product $\langle\,\cdot,\cdot\,\rangle$ is defined by
\begin{equation}
\langle\,\boldsymbol{b}(\boldsymbol{x}),\boldsymbol{c}(\boldsymbol{x})\,\rangle
=\int_{V}{\boldsymbol{b}^T(\boldsymbol{x}){\boldsymbol{c}(\boldsymbol{x})}}\,d\boldsymbol{x},
   \label{eqinner_first}
\end{equation}
where $\boldsymbol{b}(\boldsymbol{x})$ and $\boldsymbol{c}(\boldsymbol{x})$ represents vector-valued function of real numbers depend on $\boldsymbol{x}$, $V$ is the entire domain in which $\boldsymbol{x}$ is defined, and $\int_{V}\cdot\,d\boldsymbol{x}$ is to integrate over the entire domain with respect to $\boldsymbol{x}$. 

The POD is formulated by finding the orthogonal modes $\boldsymbol{\varphi}_k(\boldsymbol{x}) \in \mathcal{H}$ that satisfies the following minimization problem
\begin{equation}
\underset{\{\boldsymbol{\varphi}_k\}_{k=1}^r}  {\operatorname{argmin}} \sum^{m_\text{max}}_{m=1}{\|{\boldsymbol{q}_m(\boldsymbol{x})-\sum^r_{k=1}\langle\,\boldsymbol{q}_m(\boldsymbol{x}),\boldsymbol{\varphi}_k(\boldsymbol{x})}\,\rangle\boldsymbol{\varphi}_k(\boldsymbol{x})\|_{\boldsymbol{x}}^2},
   \label{eqminimization}
\end{equation}
where $\|\,\cdot\,\|_{\boldsymbol{x}}$ represents the distance in Hilbert space $\mathcal{H}$.
The solution to this minimization problem is determined by finding eigenvalues $\lambda_k$ and eigenvectors from 
\begin{equation}
QQ^TW_x\boldsymbol{\varphi}_k=\lambda_k\boldsymbol{\varphi}_k\,\,\,\,\,\,\,\,\,\lambda_1\geq\lambda_2\geq\cdots\geq\lambda_n,
\label{eqeigen_dis_1}
\end{equation}
where $Q$ represents a matrix of the discretized detaset $\boldsymbol{q}_m(\boldsymbol{x})$ as follows
\begin{equation}
Q =
\begin{bmatrix}
\boldsymbol{q}_1,\,\boldsymbol{q}_2,\,\cdots,\,\boldsymbol{q}_{m_{\text{max}}}
\end{bmatrix} \in \mathbb{R}^{n \times m_{\text{max}}},
\label{eq:discretizedmatrix}
\end{equation}
$W_x$  is a matrix that assigns weights to modify the discretization width when $\boldsymbol{x}$ is discretized non-uniformly, and $n$ is the number of discrete points for $\boldsymbol{x}$. 
Here, the POD modes $\boldsymbol{\varphi}_k$ are orthonormal with respect to the inner product defined by Eq. (\ref{eqinner_first}). 
The $r$ orthogonal modes $\boldsymbol{\varphi}_k(\boldsymbol{x})\,\,\,(k=1,2,3\cdots,r)$ with large eigenvalues can represent a given data $\boldsymbol{q}_m(\boldsymbol{x})$ as follows
\begin{equation}
\boldsymbol{q}_m(\boldsymbol{x})\approx\sum_{k=1}^r\,a_{k,m}\boldsymbol{\varphi}_k(\boldsymbol{x})\,\,\,(m=1,2,3,\cdots,m_{\text{max}}),
  \label{qdecom}
\end{equation}
where $a_{k,m}$ is the coupling coefficient of the orthogonal mode presented below
\begin{equation}
\begin{split}
a_{k,m}=\langle\,\boldsymbol{q}_m(\boldsymbol{x}),\boldsymbol{\varphi}_k(\boldsymbol{x})\,\rangle.
   \label{eqeigen}
\end{split}
\end{equation}


\subsection{Application of POD to Flow Fields Obtained by CFD}
The case for applying the POD to time-dependent data of velocity fields $\boldsymbol{u}(\boldsymbol{x},t)$ obtained by CFD is formulated. The time $t$ in the CFD data is assumed to be discretized into $m_\text{max}$ points uniformly, and the flow field $\boldsymbol{u}(\boldsymbol{x},t_m)\,\,\,(m=1,2,\cdots,m_{\text{max}})$ is assumed to be discretized into $n$ points by the CFD computational grid. 

The time-series data of the velocity field is divided into a time-averaged component $\boldsymbol{\bar{u}}(\boldsymbol{x})$ and a temporal fluctuating component $\boldsymbol{\tilde{u}}(\boldsymbol{x},t_m)$, presented below
\begin{equation}
\boldsymbol{u}(\boldsymbol{x},t_m)=\boldsymbol{\bar{u}}(\boldsymbol{x})+\boldsymbol{\tilde{u}}(\boldsymbol{x},t_m),
  \label{u_tildmean}
\end{equation}
where $\boldsymbol{\bar{u}}(\boldsymbol{x})$ represents
\begin{equation}
\boldsymbol{\bar{u}}(\boldsymbol{x})=\frac{1}{m_{\text{max}}}\sum^{m_\text{max}}_{m=1}\boldsymbol{u}(\boldsymbol{x},t_m).
  \label{umean}
\end{equation}
In time-series data, the energy of the time-averaged component is often much larger than that of the temporal fluctuating component\cite{mfPOD}. When the POD is applied to time-series data $\boldsymbol{{u}}(\boldsymbol{x},t_m)$, the eigenvalues of modes that capture the time-averaged component are significantly larger than the fluctuating components. Since the priority of a mode is determined by the magnitude of its eigenvalues, modes that represent the fluctuating components of the flow field are less likely to be considered major modes. However, in time series data of flow fields, capturing temporal fluctuating components, such as vortex time variation, is important. Therefore, the POD analysis is performed for the fluctuating components of the flow field. 

The matrix $Q$ in the POD is given by the fluctuating components of the velocity field as follows
\begin{equation}
Q =
\begin{bmatrix}
\boldsymbol{\tilde{u}}(\boldsymbol{x},t_1),\,\boldsymbol{\tilde{u}}(\boldsymbol{x},t_2),\,\cdots,\,\boldsymbol{\tilde{u}}(\boldsymbol{x},t_{m_{\text{max}}})
\end{bmatrix} \in \mathbb{R}^{n \times m_{\text{max}}}.
\label{eq:discretizedmatrix_u}
\end{equation}
$W_x$ is the matrix whose diagonal elements are the Jacobians at each grid cell. The time-series data of the flow field is represented by the time-averaged field and the POD modes as follows
\begin{equation}
\boldsymbol{u}(\boldsymbol{x},t_m)=\boldsymbol{\bar{u}}(\boldsymbol{x}) + \sum_{k=1}^{m_{\text{max}}}a_{k}(t_m)\boldsymbol{\varphi}_k(\boldsymbol{x}),\,\,\, (m = 1, 2, \cdots m_{\text{max}}).
  \label{cPODdecom}
\end{equation}

When a dataset is obtained from CFD, the number of spatial grid points is much larger than that of temporal snapshots. Since the size of the matrix $QQ^T$ increases depending on the number of spatial grid points $n$, the eigenvalue problem in Eq. (\ref{eqeigen_dis_1}) requires enormous memory and computation time. In this case, an alternative approach referred to as snapshot POD\cite{POD_snap} is used. The eigenvalue problem of Eq. (\ref{eqeigen_dis}) in the snapshot POD is rewritten by
\begin{equation}
Q^TW_xQ\boldsymbol{\varphi}'_k=\lambda_k\boldsymbol{\varphi}'_k,
\label{eqeigen_dis_snap}
\end{equation}
where $\boldsymbol{\varphi}'_k$ is the eigenvector of $Q^TW_xQ$ and $\lambda_k$ is the eigenvalue of $Q^TW_xQ$ which is the same eigenvalue as matrix $QQ^TW_x$. The POD mode $\boldsymbol{\varphi}_k$ is obtained by
\begin{equation}
\boldsymbol{\varphi}_k=\frac{1}{\sqrt{\lambda_k}}Q\boldsymbol{\varphi}'_k.
\label{eqeigen_dis}
\end{equation}
In this study, the POD mode is obtained by the snapshot POD.

\subsection{Global POD}
The modes obtained by the POD are the optimal modes for representing given data. However, the POD modes are not the best for representing data not contained in the dataset. Thus, the optimal mode is selected for a given condition of the flow field, but the mode is not guaranteed to represent a flow field with conditions different from those of the data. A global POD obtains a robust mode over a range of conditions rather than being optimal for a specific condition. 

The global POD is formulated for a time series of flow field data that depends on the parameter $\eta$. The flow field is represented as a discretized form $\boldsymbol{u}(\boldsymbol{x},t_m,\eta_l)\,\,\,(m=1,2,\cdots,m_{\text{max}},\,\,\,l=1,2,\cdots,l_{\text{max}})$ for $t$, $\eta$, where $t=t_1, t_2,\cdots, t_{m_{\text{max}}}$ for the conditions $\eta=\eta_1, \eta_2,\cdots, \eta_{l_{\text{max}}}$, respectively. The flow field is also discretized spatially by the computational grid.

Here, since time-series data exists at each of the discrete points in condition $\eta$, the time-averaged component can be computed for $l_{\text{max}}$ number of components by computing the time-averaged component for each condition $\eta_l$. However, when reconstructing the flow field using Eq. (\ref{cPODdecom}), multiple average components cause differences in the resulting flow field depending on which average component is used. Although no universal method exists for treating mean fields in the global POD, in this study, the mean value of all data is treated as the time-averaged component, so that the mean field used in the reconstruction is the same. Thus, the fluctuating components in the global POD are computed as follows
\begin{equation}
\boldsymbol{\tilde{u}}_g(\boldsymbol{x},t_m,\eta_l)=\boldsymbol{u}(\boldsymbol{x},t_m,\eta_l)-\boldsymbol{\bar{u}}_g(\boldsymbol{x}),
  \label{u_tildmean}
\end{equation}
where $\boldsymbol{\bar{u}}_g(\boldsymbol{x})$ represents the average component in the global POD presented below
\begin{equation}
\boldsymbol{\bar{u}}_g(\boldsymbol{x})=\frac{1}{l_{\text{max}}}\sum^{l_\text{max}}_{l=1}\left\{\frac{1}{m_{\text{max}}}\sum^{m_\text{max}}_{m=1}\boldsymbol{u}(\boldsymbol{x},t_m,\eta_l)\right\}.
  \label{umean}
\end{equation}
The global POD mode is obtained by the eigenvalue problem of Eq. (\ref{eqeigen_dis_snap}) with matrix $Q$, which is lining up $\boldsymbol{\tilde{u}}_g(\boldsymbol{x},t_m,\eta_l)$ presented below
\begin{equation}
\begin{split}
Q = [&\,\boldsymbol{\tilde{u}}_g(\boldsymbol{x},t_1,\eta_1),\,\boldsymbol{\tilde{u}}_g(\boldsymbol{x},t_2,\eta_1),\,\cdots\\
&\,\,\,\,\,\,\,\,\,\,\,\,\,\,\,\,\,\,\,\,\,\,\,\,\,\,\,\,\,\,\,\,\,\,\,\,\,\,\,\,\,\,\,\,\,\,\,\,\,\,\cdots, \,\boldsymbol{\tilde{u}}_g(\boldsymbol{x},t_{m_{\text{max}}},\eta_1),\\
&\,\boldsymbol{\tilde{u}}_g(\boldsymbol{x},t_1,\eta_2),\,\,\,\,\,\,\,\,\,\,\cdots\,\,\,\,\,\,\,\,\,\,\boldsymbol{\tilde{u}}_g(\boldsymbol{x},t_{m_{\text{max}}},\eta_2),\\
&\,\cdots\\
&\,\boldsymbol{\tilde{u}}_g(\boldsymbol{x},t_1,\eta_{l_{\text{max}}})\,\,\,\,\,\cdots,\,\,\,\,\boldsymbol{\tilde{u}}_g(\boldsymbol{x},t_{m_{\text{max}}},\eta_{l_{\text{max}}})] \\
&\,\,\,\,\,\,\,\,\,\,\,\,\,\,\,\,\,\,\,\,\,\,\,\,\,\,\,\,\,\,\,\,\,\,\,\,\,\,\,\,\,\,\,\,\,\,\,\,\,\,\,\,\,\,\,\,\,\,\,\,\in \mathbb{R}^{n \times (m_{\text{max}} l_{\text{max}})}.
\label{eq:discretizedmatrix_u}
\end{split}
\end{equation}

\subsection{Reconstruction of Flow Fields}
The eigenvalues corresponding to the POD mode are the energy of the flow field that is represented by the POD mode. Therefore, using the POD modes with large eigenvalues, the flow field is reconstructed with almost the same energy as the prepared flow data for the POD. When the number of POD modes used in the reconstruction is $r$, the flow field is approximated as follows
\begin{equation}
\boldsymbol{u}(\boldsymbol{x},t)\approx \boldsymbol{\bar{u}}(\boldsymbol{x}) + \sum_{k=1}^{r}a_{k}(t)\boldsymbol{\varphi}_k(\boldsymbol{x}).
  \label{cPODdecom}
\end{equation}

Since the eigenvalues of the POD modes correspond to the energy of the flow field that is represented by the modes, it is reasonable to choose the number of modes $r$ based on the magnitude of the eigenvalues. The decision is based on the sum of the $r$ largest eigenvalues being close to the energy of the all flow snapshot prepared for the POD. The energy of the entire flow field given in the POD corresponds to the sum of all eigenvalues, and it is formulated as the smallest $r$ satisfying the following equation
\begin{equation}
  \frac{\sum_{k=1}^{r}{\lambda}_k}{\sum_{k=1}^{m_{\text{max}}}{\lambda}_k}\geq\sigma_0,
  \label{ramdcriterion}
\end{equation}
where $\sigma_0$ is the reference value for the proportion of the total energy captured by the selected modes and depends on the literature; however, it is often used as $0.98$\cite{POD_ANN} or $0.99$\cite{SPOD}. In this paper, the number of modes $r$ is determined by $\sigma_0=0.999$ based on previous work\cite{myPoF}.

\subsection{Galerkin Projection Approach}
The coefficients $a_k$ corresponding to the POD modes need to be obtained when the flow fields are reconstructed by Eq.~(\ref{cPODdecom}). One of the methods to obtain the coefficients is based on the Galerkin projection approach\cite{POD_Galerkin}, which involves projecting the governing equations onto the low-dimensional space spanned by the POD modes. 

Assuming that the velocity $\boldsymbol{u}(\boldsymbol{x},t)$ is decomposed as shown in Eq. (\ref{cPODdecom}), the governing equations can be rewritten by
\begin{equation}
\begin{split}
\sum_{i=1}^{r}&\frac{d{a_i}}{d{t}}{{{\boldsymbol{\varphi}}_i}}\\
&=-\Bigg(\sum_{i=1}^{r}\sum_{j=1}^{r}{a_i}{a_j}\nabla\cdot\boldsymbol{\varphi}_i\boldsymbol{\varphi}_j\\
&\,\,\,\,\,\,\,\,\,\,\,+\sum_{j=1}^{r}{a_j}\nabla\cdot\boldsymbol{\bar{u}}\boldsymbol{\varphi}_j
+\sum_{i=1}^{r}{a_i}\nabla\cdot\boldsymbol{\varphi}_i\boldsymbol{\bar{u}}\Bigg)\\
&\,\,\,\,\,\,\,\,\,\,\,+\frac{1}{\text{Re}}\left({\nabla^2}\boldsymbol{\bar{u}}+\sum_{i=1}^{r}{a_i}{\nu}{\nabla^2}\boldsymbol{\varphi}_i\right)
-\frac{1}{\rho}\nabla{p}.
   \label{splitlinear}
\end{split}
\end{equation}
Since the POD mode $\boldsymbol{\varphi}_k$ is orthonormal, taking the inner product defined by Eq. (\ref{eqinner_first}) with the $k$th POD mode $\boldsymbol{\varphi}_k$,  the equation becomes
\begin{equation}
\begin{split}
\frac{d{a_k}}{d{t}}
=-\sum_{i=1}^{r}&\sum_{j=1}^{r}{a_i}{a_j}F_{ijk}\\
&+\sum_{i=1}^{r}{a_i}G_{ik}+H_{k}-{\langle\frac{1}{\rho}\nabla{p},{\boldsymbol{\varphi}}_k\rangle},
   \label{GP_withpre}
\end{split}
\end{equation}
where coefficients $F_{ijk}$, $G_{ik}$, and $H_{k}$ are respectively given by
\begin{equation}
\begin{split}
F_{ijk}&={\langle\nabla\cdot\boldsymbol{\varphi}_i\boldsymbol{\varphi}_j,{\boldsymbol{\varphi}}_k\rangle},\\
G_{ik}&={\frac{1}{\text{Re}}\langle{\nabla^2}\boldsymbol{\varphi}_i,{\boldsymbol{\varphi}}_k\rangle}
-\langle\nabla\cdot\boldsymbol{\bar{u}}\boldsymbol{\varphi}_i,{\boldsymbol{\varphi}}_k\rangle
-\langle\nabla\cdot\boldsymbol{\varphi}_i\boldsymbol{\bar{u}},{\boldsymbol{\varphi}}_k\rangle,\\
H_{k}&=\frac{1}{\text{Re}}{\langle{\nabla^2}\boldsymbol{\bar{u}},{\boldsymbol{\varphi}}_k\rangle}.
   \label{GP_withpre_sub}
\end{split}
\end{equation}
Here, under the definition of inner product Eq. (\ref{eqinner_first}), the fourth term on the right-hand side of Eq. (\ref{GP_withpre}) is negligible\cite{pressure1,pressure2}.  

Therefore, the governing equations become ordinary differential equations for $a_k$ presented below
\begin{equation}
\begin{split}
\frac{d{a_k}}{d{t}}
=-\sum_{i=1}^{r}&\sum_{j=1}^{r}{a_i}{a_j}F_{ijk}
+\sum_{i=1}^{r}{a_i}G_{ik}+H_{k}.
   \label{splitlinear}
\end{split}
\end{equation}
By time-integrating these equations, the temporal evolution of the coefficients is computed. In this study, the second-order-explicit Adams--Bashforth scheme is used for the time integration of the term $F_{ijk}$ and $H_{k}$, and the second-order-implicit \textcolor{black}{Crank--Nicolson} scheme is used for the term $G_{ik}$. The coefficients $F_{ijk}$, $G_{ik}$, and $H_{k}$ are obtained using the discretized POD modes and the average field in the general coordinate system. In computing the coefficients, the derivatives of POD modes and the average field are evaluated by the QUICK method for the nonlinear terms and by the second-order central difference scheme for the linear terms, as described in the numerical simulations. Details of time integration methods are described in Appendix \ref{apenB}. The coefficients $F_{ijk}$, $G_{ik}$, and $H_{k}$ are time-independent and computed only once during the computation, which allows for faster computations for time progress. 

\section{Application of Global POD-Based ROM for Flow Fields around Circular Cylinder}\label{globalapplication}
\subsection{Global POD modes}
\begin{figure}[!ht]
   \centering
   \includegraphics[scale=0.3]{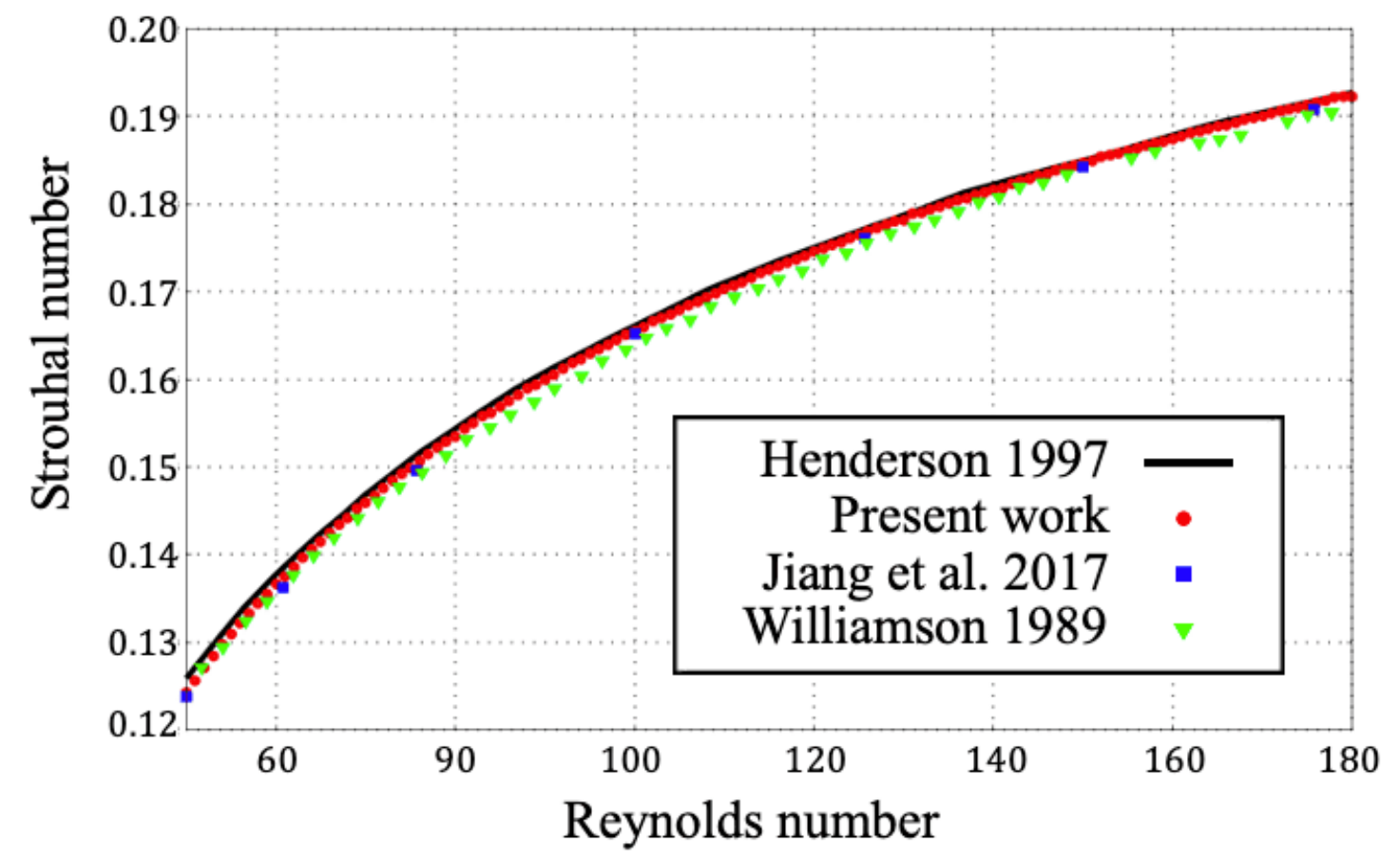}
   \caption{Reynolds number dependence of the Strouhal number for flow around a cylinder obtained by numerical simulation. The black line shows the Reynolds-Strouhal number curve in the Henderson\cite{Henderson}, the red circle is the present result of full numerical simulation, the blue square is the DNS result of Jiang et al.\cite{cylinder4}, and the green triangle is the experimental results of Williamson\cite{Williamson_1989}. }
   \label{numericalSt}
\end{figure}

An ROM using the global POD modes is constructed for flows around circular cylinders with Reynolds numbers from $50$ to $180$. The full numerical simulations were performed for $\text{Re}=50-180$ for a total of $131$ conditions, varying the Reynolds number by $1$. The Strouhal number of vortex shedding behind the cylinder obtained from numerical simulation is shown in Fig. \ref{numericalSt}. These Strouhal numbers are on the Reynolds-Strouhal number curve referred to by Henderson\cite{Henderson}.

Several conditions are selected from the $131$ Reynolds numbers, and the global POD is performed on the time series data of the flow field for the selected conditions. As shown in Table \ref{table2_1}, the global POD was performed by different numbers of conditions selected: $3$, $14$, and $27$. The values of Reynolds number were determined to be almost uniformly distributed over the range $\text{Re}=50-180$.
\begin{table*}[htbp]
  \caption{Number of conditions \textcolor{black}{included in the dataset} and Reynolds number values selected in the global POD}
  \label{table2_1}
  \centering
  \renewcommand{\arraystretch}{1.5} 
  \begin{tabular}{|c||c|}
   \hline
    number of conditions & Selected Reynolds number\\
    \hline \hline
    3   & $50$, $110$, $180$\\
    \hline
    14 & \parbox{11cm}{\centering $50$, $60$, $70$, $80$, $90$, $100$, $110$, $120$, $130$, $140$, $150$, $160$, $170$, $180$}\\
    \hline
    27 & \parbox{11cm}{\centering $50$, $55$, $60$, $65$, $70$, $75$, $80$, $85$, $90$, $95$, $100$, $105$, $110$, $115$,$120$, $125$, $130$, $135$, $140$, $145$, $150$, $155$, $160$, $165$, $170$, $175$, $180$}\\
    \hline
  \end{tabular}
\end{table*}

\begin{figure*}[hpbt]
   \centering
   \includegraphics[width=140mm]{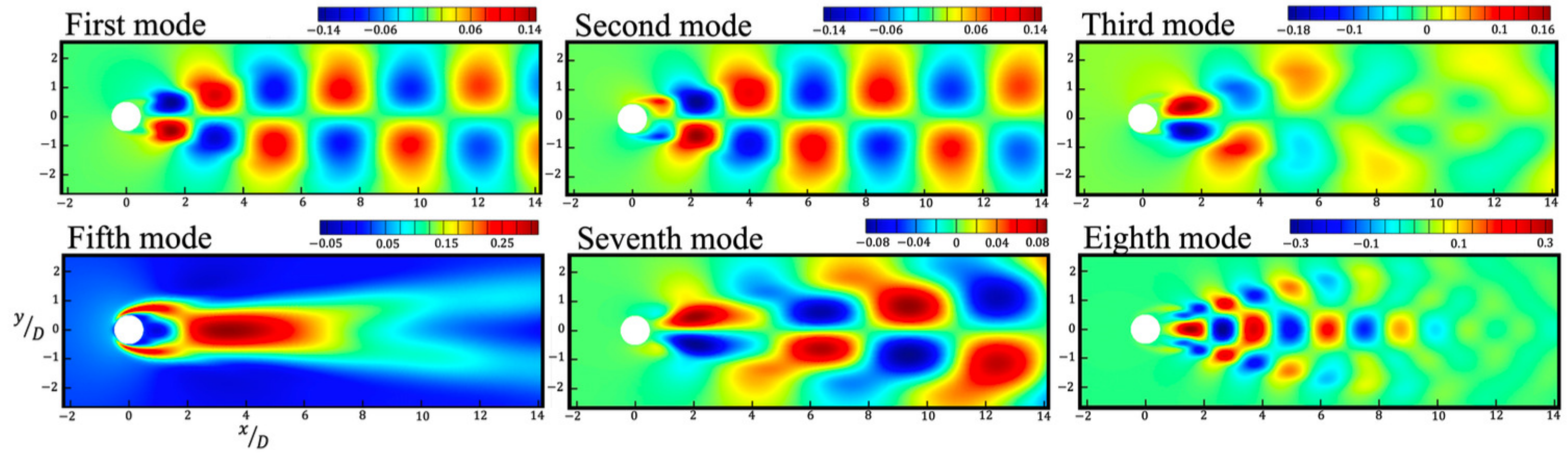}
   \caption{\textcolor{black}{Global POD modes obtained from} a dataset containing a time series flow field with three Reynolds numbers. The first and second modes form a pair. \textcolor{black}{The first and second modes form a pair. The presence of additional paired modes can be inferred from the spatial distributions of the third, seventh, and eighth modes. In contrast, the fifth mode does not form a pair with any other mode.}}
   \label{PODmode}
\end{figure*}

Figure \ref{PODmode} shows the global POD modes obtained by performing the POD on a dataset containing time series data of the flow field around a cylinder for three conditions, Reynolds numbers of $50$, $110$, and $180$. The $1$st and $2$nd POD modes have a twin spatial distribution. These two modes capture the most energetically vortex structure in the cylinder wake, with vortices distributed as far back as $x/D=14$. The $3$rd POD mode also has a vortex structure with a spatial scale similar to that of the $1$st and $2$nd modes, but it has a distribution only up to $x/D=6$ behind the cylinder. In addition, the vortices have almost the same spatial scale as the $1$st and $2$nd modes up to the $7$th mode, and the modes larger than the $8$th mode capture smaller vortices. Thus, there are three pairs of modes with large vortex structures between $1$st and $7$th. The number of pairs matches the number of conditions in the dataset. Since the position and size of the vortex near the cylinder differ with different Reynolds numbers of the flow field, the $3$rd and $7$th modes capture a large vortex structure not captured by the $1$st and $2$nd modes. 

The $5$th mode has no pairing with either mode. This mode's spatial distribution is similar to the shift mode introduced by Noack et al.\cite{shift}. The shift mode captures the difference between the size of the recirculation region in the steady Navier--Stokes equation and that of the mean flow obtained by averaging the quasi-steady flow field. In the present case, the steady Navier--Stokes solution is not included in the dataset. However, the size of the recirculation region depends on the Reynolds number of the flow field, so the dataset of the global POD contains flow fields with different sizes of the recirculation region. Therefore, this $5$th mode is regarded as a mode due to the difference in the recirculation region size originating from the Reynolds number difference.

\subsection{Results of Global POD-Based ROM}
\begin{figure}[!ht]
   \centering
   \includegraphics[width=8.0cm]{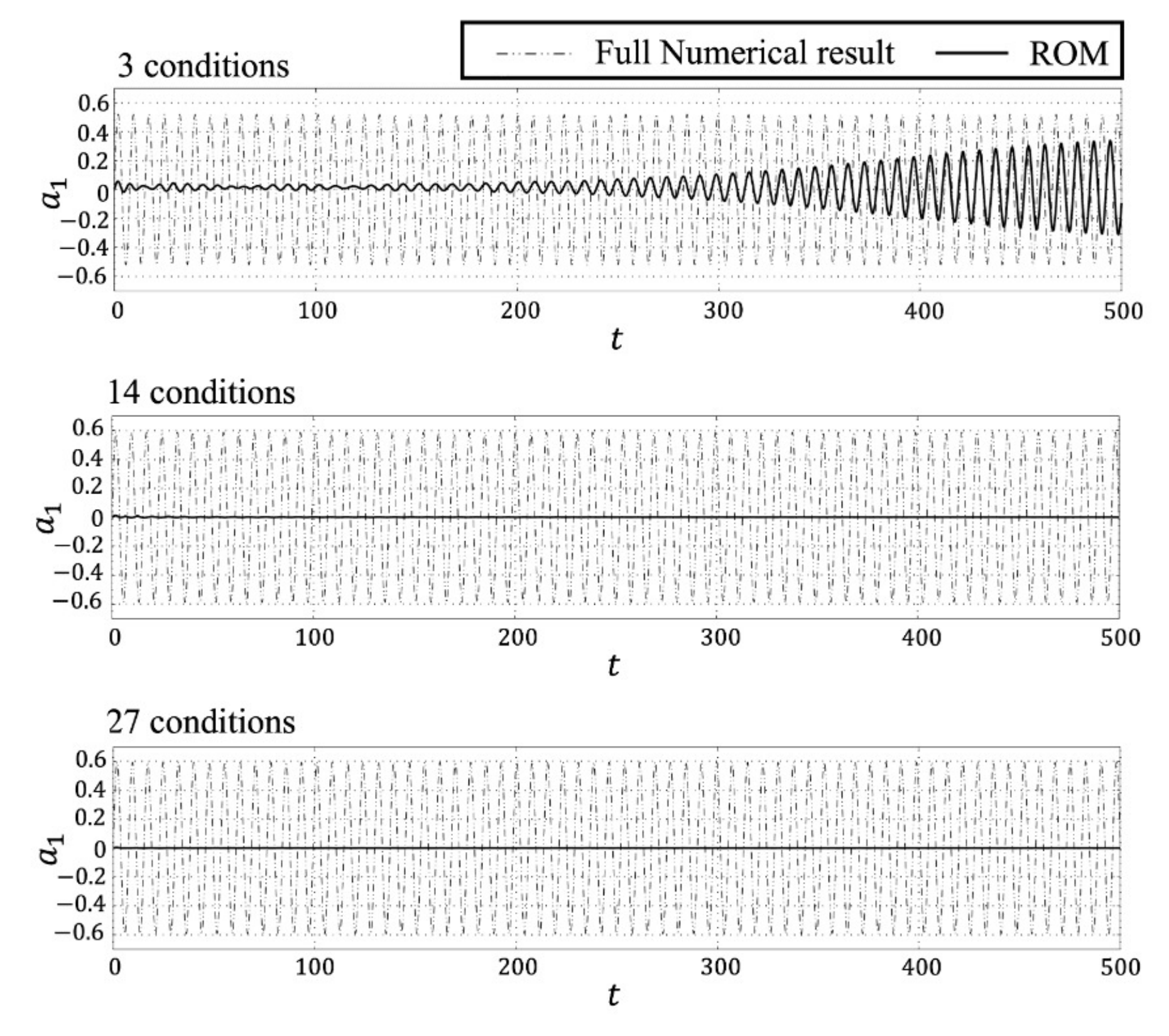}
   \caption{Time variation of global POD mode coefficients at Reynolds number $55$ when ROM is constructed by changing the number of conditions in the dataset. The initial values for the coefficients are set to $0$. Full numerical results in the figure are computed from projecting the flow field into POD modes.}
   \label{coefficient_glPOD_55}
\end{figure}

ROMs are constructed by the global POD modes obtained from $3$, $14$, and $27$ conditions, respectively. The time evolution of the coefficients obtained by the governing equation given by the Galerkin projection approach at Reynolds number $55$ is shown in Fig. \ref{coefficient_glPOD_55}. The initial condition for each coefficient is set to $0$. The ROM, constructed using the flow fields with three conditions, undergoes periodic oscillations at Reynolds number $55$ when computed for a sufficient amount of time. The ROM constructed using the flow field with $14$ or $27$ conditions shows no periodic oscillations, resulting in a steady-state solution. The ROM constructed by the flow field of $27$ and $14$ conditions fails to reproduce the flow field of Reynolds number $55$ since a full numerical simulation of the flow field at the same Reynolds number results in a periodic flow field. Therefore, although the flow features that can be represented by the POD mode have increased due to the increase in the number of conditions included in the dataset, a solution that differs from the given data is obtained.

\begin{figure}[!ht]
   \centering
   \includegraphics[width=8.0cm]{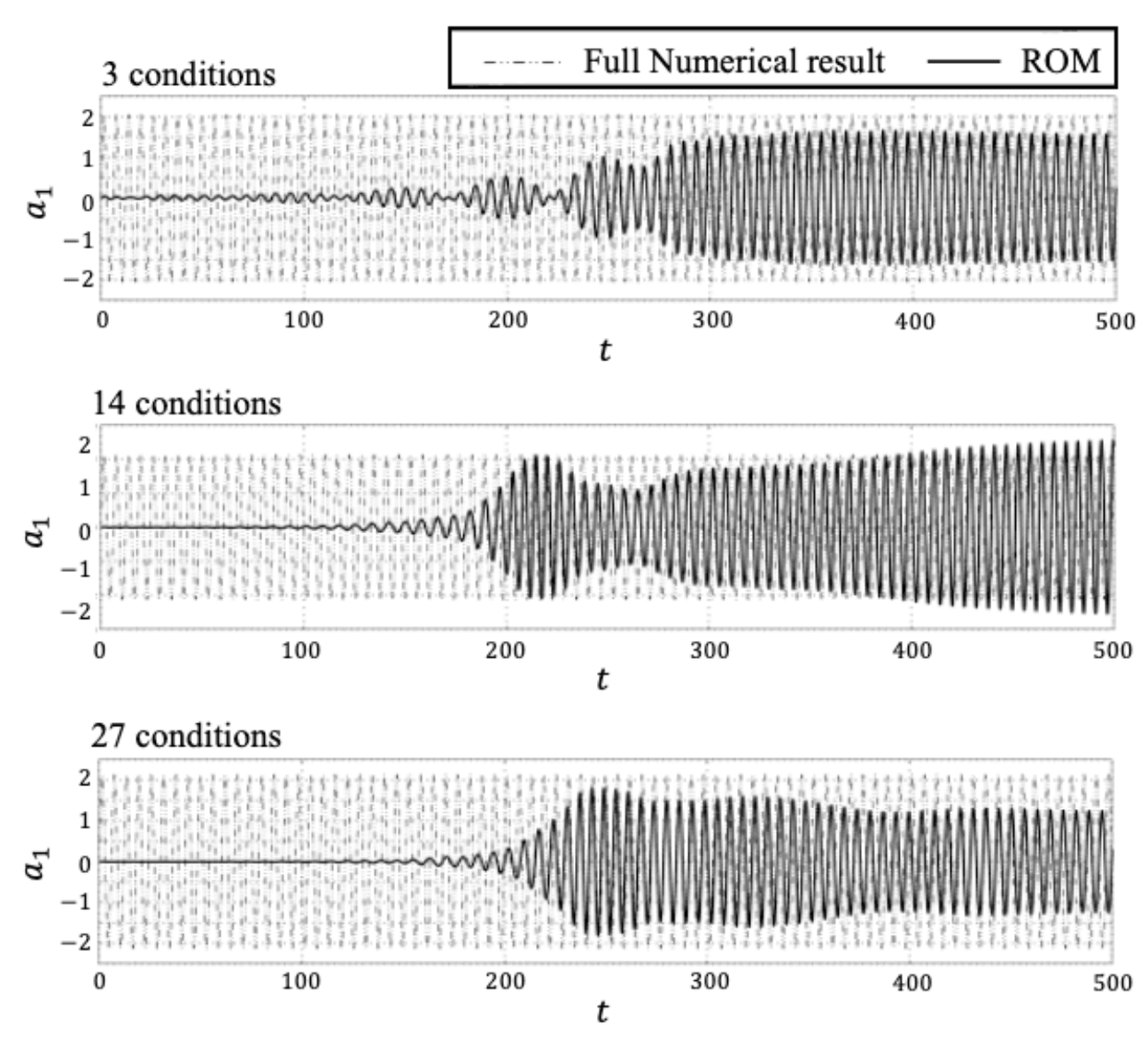}
   \caption{Time variation of POD mode coefficients at Reynolds number $100$ when ROM is constructed by changing the number of conditions in the dataset in global POD. }
   \label{coefficient_glPOD_100}
\end{figure}

The time evolution of the coefficients at Reynolds number $100$ is shown in Fig. \ref{coefficient_glPOD_100}.  At a Reynolds number of $100$, the ROM results show convergence to a solution with periodic oscillations after a sufficient time evolution. This is similar to the results of full-order numerical simulation. However, the oscillation amplitude of the coefficients differs between the results obtained by projecting the flow field obtained by full numerical simulation into the POD mode and the solution obtained in the ROM. It also shows that the more conditions used in constructing the ROM, not all of the results of the ROM are close to the results of the full numerical simulation.

\begin{figure}[!ht]
   \centering
   \includegraphics[width=8.0cm]{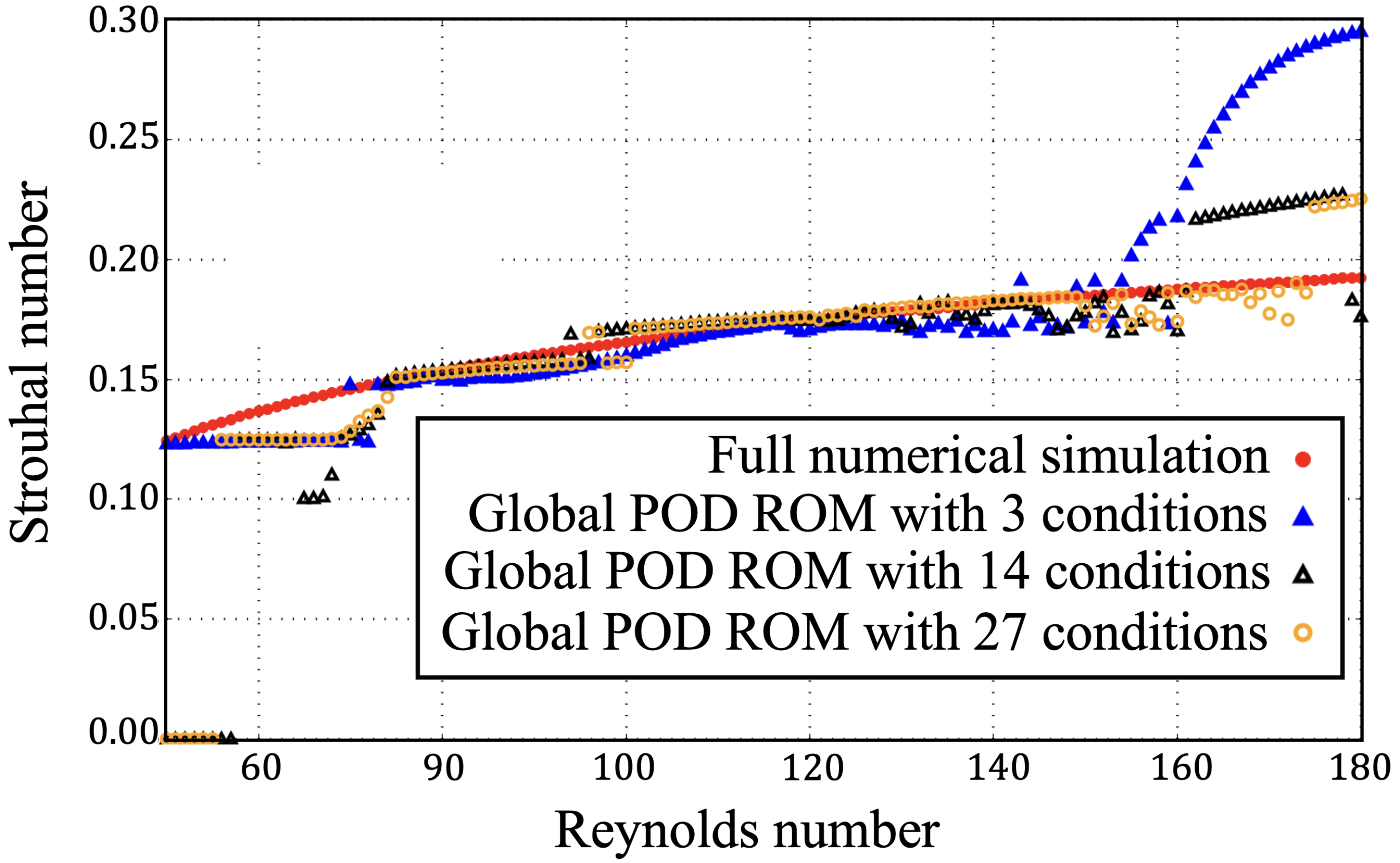}
   \caption{Strouhal number with varying Reynolds number obtained by full numerical simulation and ROMs. The three ROMs are constructed by a set of POD modes obtained by performing global POD on flow fields \textcolor{black}{at $3$, $14$, and $27$ different Reynolds numbers, respectively.}}
   \label{coefficient_St_glPOD}
\end{figure}

Next, the ROM is performed by varying the Reynolds number by $1$ in the range of Reynolds numbers from $50$ to $180$. For each Reynolds number, the mode with the largest amplitude is selected for the coefficients that reached a quasi-steady or stable state after a sufficiently long time evolution. The non-dimensionalized frequencies for coefficient oscillation are shown in Fig. \ref{coefficient_St_glPOD}. The critical Reynolds number between the quasi-steady flow and the steady flow depends on the number of conditions contained in the dataset. In the full numerical simulations, $\text{Re} \approx 46$ is the bifurcation point between quasi-steady and steady flow\cite{cylinderreview}. Hence, ROMs other than those made from datasets containing the three conditions indicate the incorrect bifurcation point. The larger critical Reynolds number in ROM compared to full-order numerical simulation is consistent with the results of Noack et al.\cite{shift}. For $\text{Re}>160$, Strouhal numbers obtained by ROMs are closer to the full numerical results as the number of conditions in the dataset increases. Therefore, to improve the robustness of ROM, it is effective to include flow fields with many conditions in the dataset.

The time required to compute the flow field is compared between each ROM. Table \ref{table4_2} shows the average CPU time required to compute the $131$ conditions for each ROM. The CPU time is the time required to compute 50,000 time steps in the Galerkin projection approach. The number of time steps is determined based on the time it takes for the solution to converge in ROM. Computations were performed using an Intel Xeon Gold 6148 2.4GHz×2 CPU on the Supercomputer system ``AFI-NITY" at the Advanced Fluid Information Research Center, Institute of Fluid Science, Tohoku University. The ROM computation cost increases as the number of the dataset's conditions increases. The increased number of conditions in the dataset increases the number of POD modes used in the ROM, resulting in increased computation time. Another possible factor is the worsening convergence of the iterative method in the implicit time-integral method in the Galerkin projection approach.

\begin{table}[tbp]
  \caption{Comparison of CPU time required to predict flow field \textcolor{black}{by the ROMs}. Computational cost increases as the number of conditions contained in the dataset increases.}
  \label{table4_2}
  \centering
  \renewcommand{\arraystretch}{1} 
  \begin{tabular}{|c||c|}
   \hline
    Number of conditions & CPU time\\
    \hline \hline
    3  & $83$\\
    \hline
    14 & $134$\\
    \hline
    27 & $206$\\
    \hline
  \end{tabular}
\end{table}


\section{Dual-Step POD-Based ROM}\label{dualmethod}
\subsection{Concept}
In the conventional global POD-based ROM, as the number of conditions given to the POD dataset increases, the computational cost required to compute the flow field by ROM increases. However, the number of flow conditions required to reconstruct a flow field at a given Reynolds number is significantly less than that contained in the datasets for the global POD. Therefore, reducing the number of flow conditions required for reconstruction is possible when the POD is performed on only flow fields that are suitable for the Reynolds number to be reconstructed.
 
\begin{figure}[!ht]
   \centering
   \includegraphics[width=8.0cm]{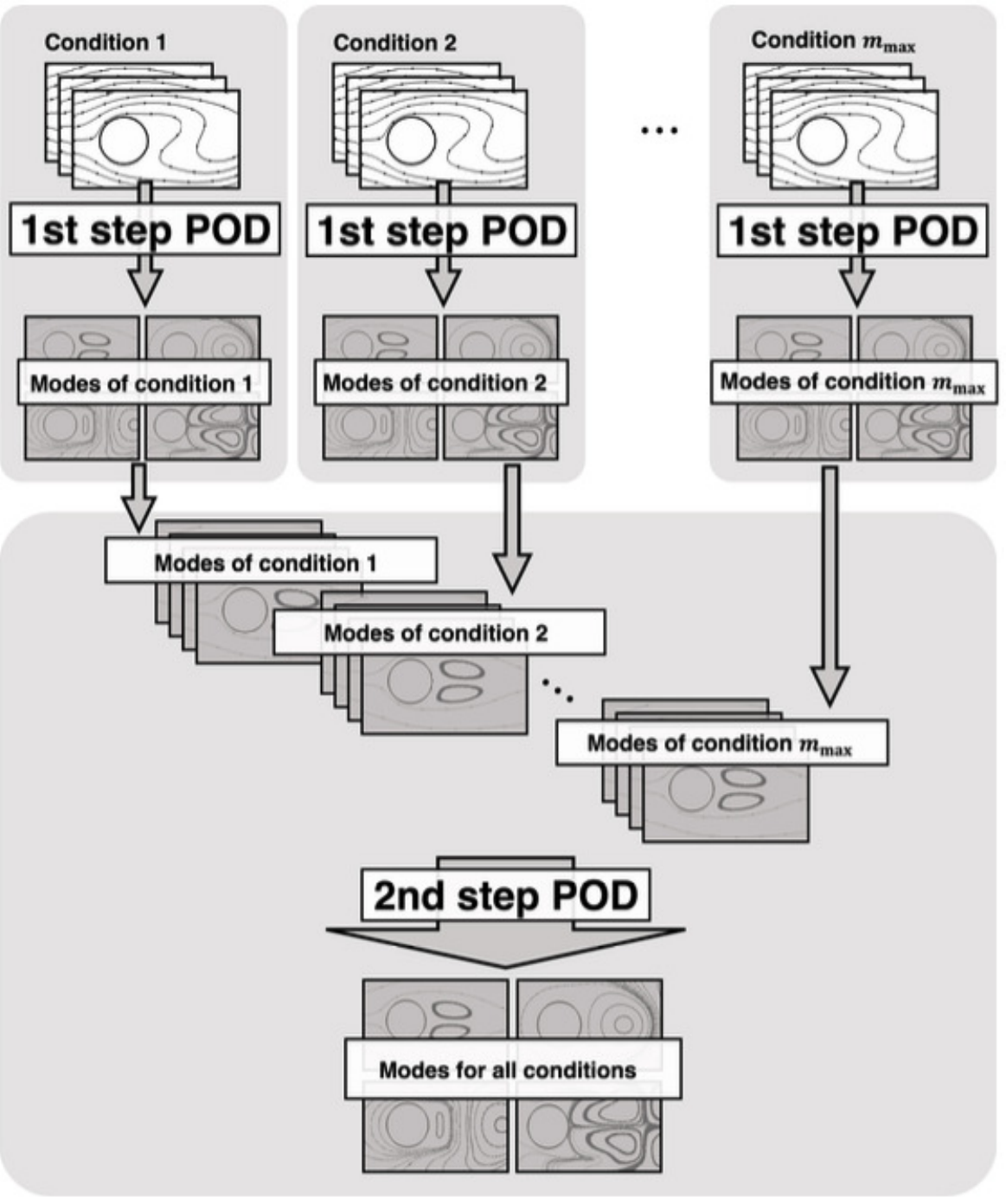}
   \caption{The procedure of dual-step POD. In the first step, POD is performed on each condition individually; in the second step, POD is performed on the mode of all conditions to obtain a set of global modes.}
   \label{dualPOD}
\end{figure}

We propose an ROM based on dual-step POD for data containing flow fields under various Reynolds numbers. Figure \ref{dualPOD} shows the dual-step POD procedure for data containing multiple conditions. In the first step of dual-step POD, the POD is performed for flow snapshots at each Reynolds number to obtain the optimal POD mode for the reference condition. Hence, the set of POD modes can be as many as the number of conditions contained in the entire dataset. In the second step, the POD is performed on the set of all modes obtained in the first step. The second step results in a set of POD modes for representing the entire dataset, similar to a conventional global POD. 

\begin{figure}[hpbt]
   \centering
   \includegraphics[width=8.0cm]{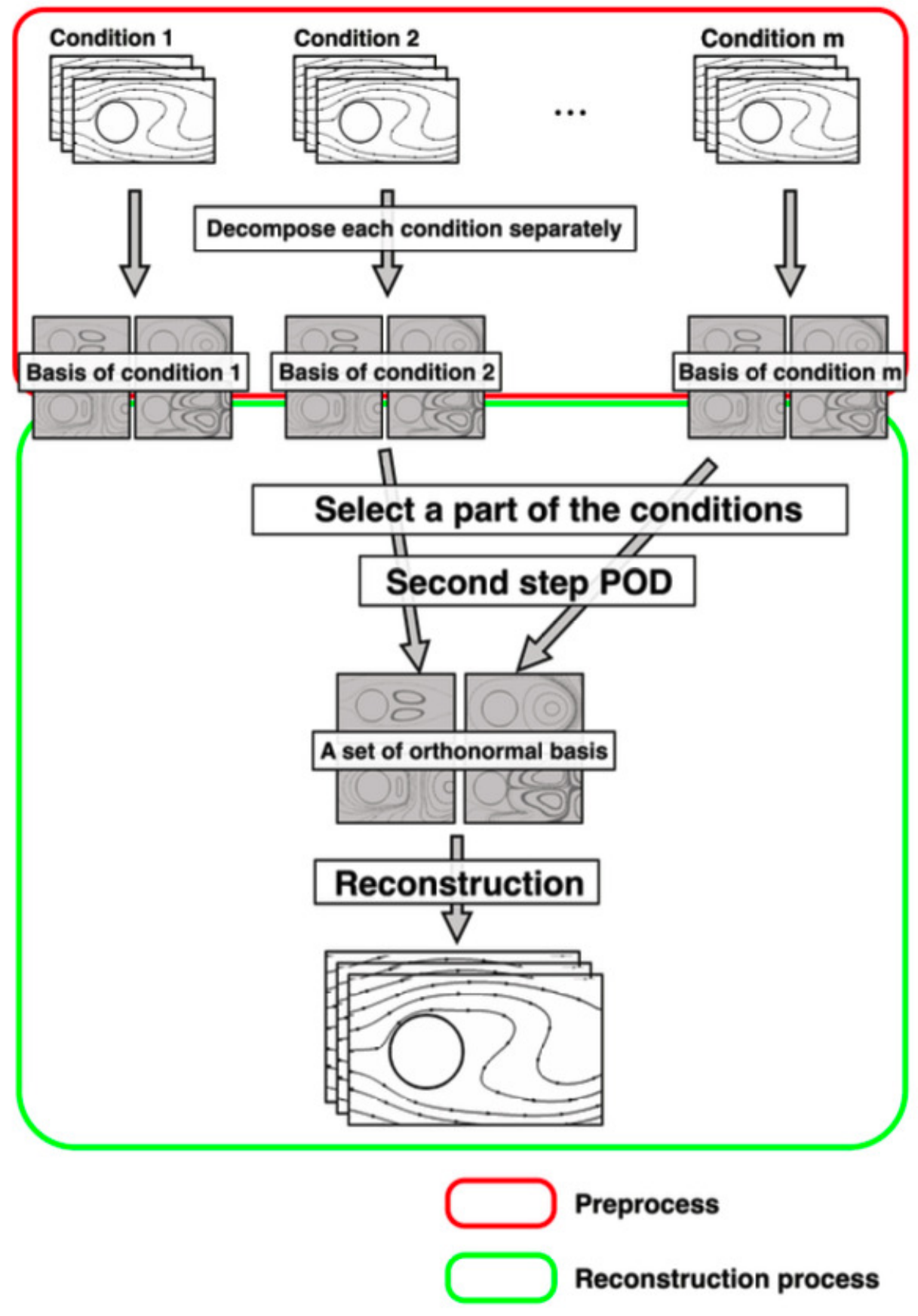}
   \caption{The structure of the proposed ROM based on dual-step POD. The main idea is to include the second step of \textcolor{black}{dual-step} POD in the ROM computation process. By performing the second step only on POD modes that are close to the reconstruction conditions of the ROM, the optimal POD modes for the reconstruction conditions are obtained.}
   \label{dualROM}
\end{figure}

Sato et al.\cite{Sato_prf} proposed to connect POD modes with individual conditions by applying POD to these POD modes. It is shown that the POD mode obtained from the POD mode can represent the solution for different conditions from the dataset in the Burgers equation. In this paper, we select the POD modes with features close to the target Reynolds number to be represented instead of using the POD mode for the entire dataset.  Thus, a part of all the sets of first-step POD modes is selected in the second-step POD. \textcolor{black}{The structure of proposed ROM using the dual-step POD is} shown in Fig. \ref{dualROM}.  An ROM using dual-step POD is constructed using the first step POD modes of the dual-step POD process, maintaining multiple sets of optimal POD modes for each condition. After the flow conditions to be predicted in ROM are settled, the second-step POD is performed by selecting only the flow field with conditions close to those conditions. Thus, the second step POD \textcolor{black}{process} is part of the ROM computation process. The computation of the second step POD does not \textcolor{black}{significantly} increase the ROM computation cost since the second POD is performed only for the POD modes, and this computation takes little time.

\subsection{Methodology of mfPOD}
In the dual-step POD, there is a critical issue of how to handle the mean field. Since the mean field is typically the average of the entire dataset for which the POD is performed, the first POD in the dual-step POD computes as many mean fields as the condition. In this case, the accuracy of the prediction depends on which mean field is used for the reconstruction of the ROM. To provide uniform treatment of mean fields, we use $mf$POD\cite{mfPOD}, which includes mean fields in the datasets.

The $mf$POD is a technique that extracts features of the datasets, including the average of the datasets.  The matrix $Q$ in the eigenvalue problem of the $mf$POD is created by lining up the $m$ fluctuating components $\boldsymbol{\tilde{u}}_1, \boldsymbol{\tilde{u}}_2, \cdots , \boldsymbol{\tilde{u}}_{m_{\max}}$  and the average-field $\boldsymbol{\bar{u}}$ as follows
 \begin{equation}
U=[\boldsymbol{\tilde{u}}(\boldsymbol{x},t_1), \boldsymbol{\tilde{u}}(\boldsymbol{x},t_2), \cdots , \boldsymbol{\tilde{u}}(\boldsymbol{x},t_{m_{\text{max}}}), \sqrt{\omega}\boldsymbol{\bar{u}}(\boldsymbol{x})],
  \label{mf1}
\end{equation}
where $\omega$ is a scalar weight. In the $mf$POD, the mean-field energy is weighted, and this scalar weight controls the mean-field energy.  The larger $\omega$ is, the larger the mean-field energy considered when selecting the modes. Decomposing $U$ into $r$ modes by the POD, $\boldsymbol{\tilde{u}}_1, \boldsymbol{\tilde{u}}_2, \cdots , \boldsymbol{\tilde{u}}_{m_{\text{max}}}$ and $\boldsymbol{\bar{u}}$ are expressed as follows
\begin{equation}
\boldsymbol{\tilde{u}}(\boldsymbol{x},t) \approx \sum_{k=1}^{r}\tilde{a}_{k}(t)\boldsymbol{\psi}_k(\boldsymbol{x}).
  \label{mf2}
\end{equation}
\begin{equation}
\boldsymbol{\bar{u}}(\boldsymbol{x}) \approx \sum_{k=1}^{r}\bar{a}_{k}\boldsymbol{\psi}_k(\boldsymbol{x}).
  \label{mf3}
\end{equation}
From Eq. (\ref{mf2}) and Eq. (\ref{mf3}), the velocity field is decomposed in the following
\begin{equation}
\boldsymbol{u}(\boldsymbol{x},t)=\boldsymbol{\tilde{u}}(\boldsymbol{x},t) + \boldsymbol{\bar{u}}(\boldsymbol{x}) \approx \sum_{k=1}^{r}(\tilde{a}_{k}(t)+\bar{a}_{k})\boldsymbol{\psi}_k(\boldsymbol{x}).
  \label{mf4}
\end{equation}

Based on a previous study\cite{mfPOD}, the number of modes r is determined via the scalar parameter $\epsilon$ presented below
\begin{equation}
\begin{split}
\frac{\sum_{k=1}^{r}\lambda_k}{\sum_{k=1}^{m_{\text{max}}}\lambda_k}\geq\frac{\epsilon+\sigma_0}{\epsilon+1},
\label{mf13}
\end{split}
\end{equation}
where $\epsilon$ is computed as follows
\begin{equation}
  \epsilon={\omega}\frac{|\boldsymbol{\bar{u}}(\boldsymbol{x})|^2}{\sum_{j=1}^{m_{\text{max}}}|\boldsymbol{\tilde{u}}(\boldsymbol{x},t_j)|^2}.
  \label{mfcriterion}
\end{equation}
The parameter $\epsilon$ corresponds to the ratio of the energy of the mean field to that of the fluctuating components in the sum of the eigenvalues of the POD mode. The value of $\epsilon$ must be sufficiently large to represent the mean field of the flow field by the POD modes. In this study, $\omega$ is set so that $\epsilon = 10000$.

\subsection{Methodology of Dual-Step POD}
The dual-step POD is performed when the dataset contains time series flow data for two or more conditions. The flow conditions are assumed to be expressed as the parameter $\eta$. The flow field is represented as a discretized form $\boldsymbol{u}(\boldsymbol{x},t_m,\eta_l)\,\,\,(m=1,2,\cdots,m_{\text{max}},\,\,\,l=1,2,\cdots,l_{\text{max}})$ for $t$, $\eta$, where $t=t_1, t_2,\cdots, t_{m_{\text{max}}}$ for the conditions $\eta=\eta_1, \eta_2,\cdots, \eta_{l_{\text{max}}}$, respectively. Note that the flow field set $\boldsymbol{u}(\boldsymbol{x},t_m,\eta_l)$ here is the same form as the dataset used in the global POD. The first step of the dual-step POD performs $mf$POD for a specific condition $\eta_l$. Hence, the POD mode of the first step $\boldsymbol{\psi}^{1st}_k(\boldsymbol{x},\eta_l)$ is presented below
\begin{equation}
({U^{1st}_l})^TW_x{U^{1st}_l}\boldsymbol{\psi}'^{1st}_k(\eta_l)={\lambda^{1st}_k(\eta_l)}\boldsymbol{\psi}'^{1st}_k(\eta_l),
\label{eqeigen_dis_snap_dual}
\end{equation}
\begin{equation}
\boldsymbol{\psi}^{1st}_k(\boldsymbol{x},\eta_l)={U^{1st}_l}\boldsymbol{\psi}'^{1st}_k(\eta_l).
\label{eqeigen_dis_dual_first}
\end{equation}
where matrix $U^{1st}_l$ represent
 \begin{equation}
  \begin{split}
U^{1st}_l&=[\boldsymbol{\tilde{u}}(\boldsymbol{x},t_1,\eta_l), \boldsymbol{\tilde{u}}(\boldsymbol{x},t_2,\eta_l), \cdots\\
 &\,\,\,\, \cdots,\,\boldsymbol{\tilde{u}}(\boldsymbol{x},t_{m_{\text{max}}},\eta_l), \sqrt{\omega_l}\boldsymbol{\bar{u}}(\boldsymbol{x},\eta_l)],
  \label{dual1}
   \end{split}
\end{equation}
$\boldsymbol{\tilde{u}}(\boldsymbol{x},t_m,\eta_l)$ and $\boldsymbol{\bar{u}}(\boldsymbol{x},\eta_l)$ are
\begin{equation}
\boldsymbol{\tilde{u}}(\boldsymbol{x},t_m,\eta_l)=\boldsymbol{u}(\boldsymbol{x},t_m,\eta_l)-\boldsymbol{\bar{u}}(\boldsymbol{x},\eta_l),
  \label{u_tildmean}
\end{equation}
\begin{equation}
\boldsymbol{\bar{u}}(\boldsymbol{x},\eta_l)=\frac{1}{m_{\text{max}}}\sum^{m_\text{max}}_{m=1}\boldsymbol{u}(\boldsymbol{x},t_m,\eta_l),
  \label{umean}
\end{equation}
${\omega_l}$ is the weight parameter in condition $\eta_l$, determined by the scalar value $\epsilon$ common to all conditions as follows
\begin{equation}
  {\omega_l}=\epsilon\frac{\sum_{j=1}^{m_{\text{max}}}|\boldsymbol{\tilde{u}}(\boldsymbol{x},t_j,\eta_l)|^2}{|\boldsymbol{\bar{u}}(\boldsymbol{x},\eta_l)|^2}.
  \label{mfcriterion_dual}
\end{equation}
Note that the first step POD mode $\boldsymbol{\psi}^{1st}_k(\boldsymbol{x},\eta_l)$ is not normalized in eq. (\ref{eqeigen_dis_dual_first}) so that the magnitude of the physical energy captured by the POD mode corresponds to the norm of the POD mode.

In the second-step POD, POD is performed in the POD modes. Therefore, the POD mode of the second step can be obtained by following the eigenvalue problem
\begin{equation}
({U^{2nd}})^TW_x{U^{2nd}_l}\boldsymbol{\psi}'^{2nd}_k={\lambda_k}^{2nd}\boldsymbol{\psi}'^{2nd}_k,
\label{eqeigen_dis_snap_dual}
\end{equation}
\begin{equation}
\boldsymbol{\psi}^{2nd}_k(\boldsymbol{x})=\frac{1}{\sqrt{\lambda_k^{2nd}}}{U^{2nd}}\boldsymbol{\psi}'^{2nd}_k.
\label{eqeigen_dis_dual}
\end{equation}
where matrix $U^{2nd}$ represent
 \begin{equation}
  \begin{split}
U^{2nd}= [\boldsymbol{\psi}^{1st}_1(\boldsymbol{x},\eta_{1}), \boldsymbol{\psi}^{1st}_2(\boldsymbol{x},\eta_{1}), \cdots , \boldsymbol{\psi}^{1st}_{r^1}(\boldsymbol{x},\eta_{1}),\,\,\,\\
 \boldsymbol{\psi}^{1st}_1(\boldsymbol{x},\eta_{2}), \cdots , \boldsymbol{\psi}^{1st}_{r^2}(\boldsymbol{x},\eta_{2}),\,\,\,\\
  \cdots , \boldsymbol{\psi}^{1st}_{r^m}(\boldsymbol{x},\eta_{l_\text{max}}) ],
  \label{dual1}
  \end{split}
\end{equation}
and the number of POD modes $r_l$ in the first step is determined for each condition $\eta_{l}$. In this paper, the number of modes $r_l$ is the smallest $r$ that satisfies the following equation
\begin{equation}
\begin{split}
\frac{\sum_{k=1}^{r_l}\lambda^{1st}_k(\eta_{l})}{\sum_{k=1}^{m_{\text{max}}}\lambda^{1st}_k(\eta_{l})}\geq\frac{\epsilon+\sigma_0}{\epsilon+1},
\label{mf13}
\end{split}
\end{equation}
where $\sigma_0$ is set to $0.999$.

\subsection{Galerkin Projection Approach with Dual-Step POD Modes}
Even when using the POD mode obtained by the dual-step POD, the reduced governing equation can be derived as follows
\begin{equation}
\begin{split}
\frac{d{a_k}}{d{t}}
=-\sum_{i=1}^{r}&\sum_{j=1}^{r}{a_i}{a_j}F_{ijk}
+\sum_{i=1}^{r}{a_i}G_{ik}-{\langle\frac{1}{\rho}\nabla{p},{\boldsymbol{\psi}}^{2nd}_k\rangle},\\
   \label{GP_withpre_dual}
\end{split}
\end{equation}
where coefficient $F_{ijk}$ and $G_{ik}$ are
\begin{equation}
\begin{split}
F_{ijk}&={\langle\nabla\cdot\boldsymbol{\psi}^{2nd}_i\boldsymbol{\psi}^{2nd}_j,{\boldsymbol{\psi}}^{2nd}_k\rangle},\\
G_{ik}&={\frac{1}{\text{Re}}\langle{\nabla^2}\boldsymbol{\psi}^{2nd}_i,{\boldsymbol{\psi}}^{2nd}_k\rangle}.
   \label{GP_withpre_sub_dual}
\end{split}
\end{equation}
In the dual-step POD, the inner product of the third term with the mode representing the mean field is not zero since the POD mode representing the mean field is contained in ${\boldsymbol{\psi}}^{2nd}_k (k=1,2,\cdots,r)$. However, the mean field has more energy than the fluctuating components, so the mode with the largest eigenvalue corresponds to the mean field. At the inflow boundary, the boundary condition $\boldsymbol{u}=\boldsymbol{U_{\infty}}$ is imposed at all time steps, so the following holds
\begin{equation}
\boldsymbol{\psi}_i^{2nd}(x){|_{x=\Omega_{\text{inlet}}}}=\delta_{i1}\frac{\boldsymbol{U}_{\infty}}{a_i}.
   \label{eqinner_2nd}
\end{equation}
where $\delta_{ij}$ represents the Kronecker delta. Therefore, for terms other than $\boldsymbol{\psi}^{2nd}_1$, the pressure term can be neglected as in the Galerkin projection using the conventional POD modes. 

The following equation must hold at the boundary for $\boldsymbol{\psi}^{2nd}_1$
\begin{equation}
\begin{split}
a_1\boldsymbol{\psi}^{2nd}_1{|_{x=\Omega_{\text{inlet}}}} = \boldsymbol{U_{\infty}}.
\label{inletboundary_dualPOD}
\end{split}
\end{equation}
Hence, by integrating on the inflow boundary, $a_1$ can be obtained below
\begin{equation}
\begin{split}
a_1 = \frac{\int_{\Omega_{inlet}} \boldsymbol{U_{\infty}} \cdot \boldsymbol{\psi}^{2nd}_1 \, d\Omega}{\int_{\Omega_{inlet}} \boldsymbol{\psi}^{2nd}_1 \cdot \boldsymbol{\psi}^{2nd}_1 \, d\Omega}.
\label{inletboundary_dualPOD}
\end{split}
\end{equation}
Thus, the time evolution of the coefficients in the proposed ROM can be summarized as follows
\begin{equation}
\begin{split}
a_k &= \frac{\int_{\Omega_{inlet}} \boldsymbol{U_{\infty}} \cdot \boldsymbol{\psi}^{2nd}_1 \, d\Omega}{\int_{\Omega_{inlet}} \boldsymbol{\psi}^{2nd}_1 \cdot \boldsymbol{\psi}^{2nd}_1 \, d\Omega}\,\,\,\,\,\,\,\,\,\,\,\,\,\,\,\,\,\,\,\,\,\,\,\,\,(k=1),\\
\frac{d{a_k}}{d{t}}
&=-\sum_{i=1}^{r}\sum_{j=1}^{r}{a_i}{a_j}F_{ijk}
+\sum_{i=1}^{r}{a_i}G_{ik}\,\,\,(k\neq1).\\
   \label{GP_nopre_dual}
\end{split}
\end{equation}

\section{Comparison with Global POD-Based ROMs}\label{dualapplication}
An ROM using the dual-step POD is constructed using the flow field for the $27$\textcolor{black}{, $14$, and $3$} conditions in Table \ref{table2_1}. In the first step of the dual-step POD, $mf$POD was performed \textcolor{black}{on the time-series data for each condition.} The second-step POD is performed on the set of POD modes obtained from the flow field with the Reynolds number closest to the Reynolds number of the flow field to be predicted. In this study, the conditions with the closest values, one equal to or larger and one smaller than the Reynolds number of the predicted flow field, are selected in the second step of dual-step POD. 

\begin{figure}[!ht]
   \centering
   \includegraphics[width=10.0cm]{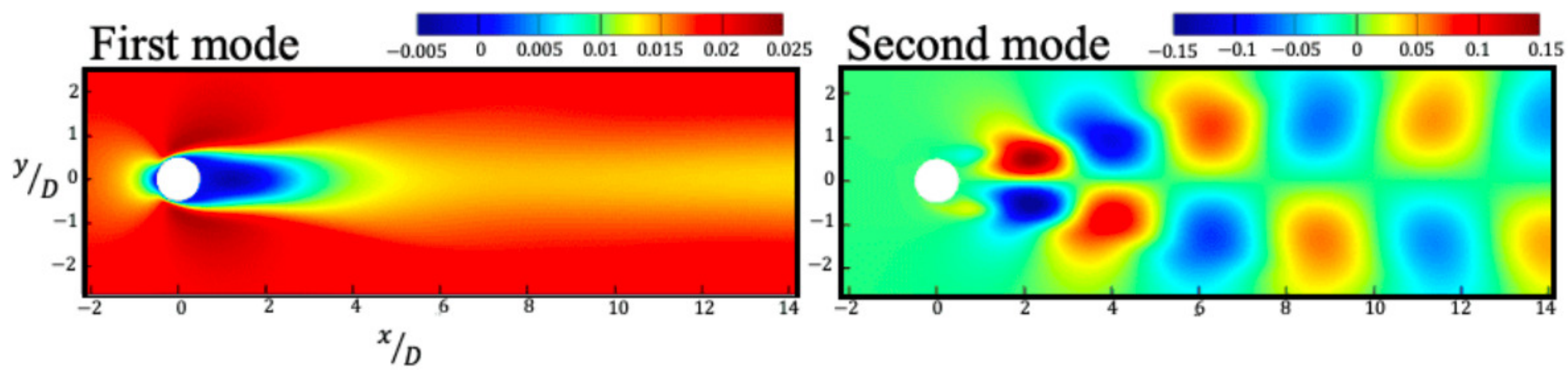}
   \caption{Modes obtained by the dual-step POD for the predicted Reynolds number of $100$. The first mode represents the mean field. The presence of paired modes is inferred from the spatial distribution of the 2nd mode.}
   \label{PODmode_dualPOD}
\end{figure}

The time evolution of the coefficients in the POD mode is computed for an ROM using the dual-step POD with a Reynolds number of $100$. The POD mode obtained with the dual-step POD to predict a Reynolds number of $100$ is shown in Fig. \ref{PODmode_dualPOD}. Since the mean field is not subtracted from the dataset in the dual-step POD, the 1st mode has a spatial distribution corresponding to the mean field. The 2nd mode captures the spatial distribution of the vortex in the wake of the cylinder and is similar to the 1st mode of the conventional global POD.

\begin{figure}[!ht]
   \centering
   \includegraphics[width=16.0cm]{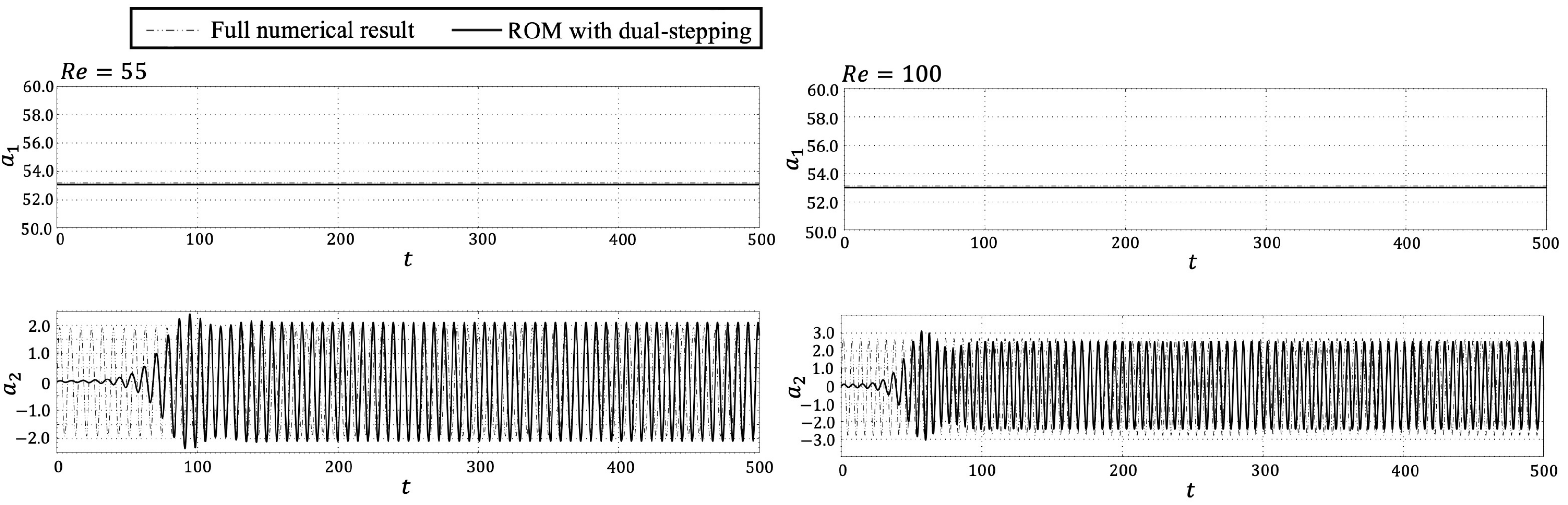}
   \caption{Time variation of POD mode coefficients at Reynolds numbers of \textcolor{black}{$55$ and} $100$. The ROM is based on dual-step POD \textcolor{black}{constructed with a dataset including 27 conditions}. The initial values for the coefficients, except for the first mode, are set to $0$. Full numerical results in the figure are computed by projecting the flow field into POD mode.}
   \label{coefficient_long_dualPOD}
\end{figure}

Figure \ref{coefficient_long_dualPOD} shows the time evolution of the coefficients for the first and second-step POD modes. For the computation of the Reynolds number $100$, a second-step POD is performed by selecting a set of POD modes for the two conditions of Reynolds numbers $100$ and $95$. \textcolor{black}{Similarly, for the computation at Reynolds number $55$, the second-step POD is conducted using POD modes selected from the datasets at Reynolds numbers $55$ and $60$.} Equation (\ref{GP_nopre_dual}) is computed for $10$ POD modes based on $\sigma_0=0.999$ for the POD mode obtained in the second-step POD. The first mode is fixed at a value that satisfies the inflow condition, as shown in Eq. (\ref{GP_nopre_dual}). Hence, the coefficients of the first mode are constant at all time steps. The second mode starts with an initial value of $0$ and converges to an almost quasi-steady solution after about $100$ in dimensionless time. The ROM based on the dual-step POD converges in less than half the time than the ROM based on the global POD shown in Fig. \ref{coefficient_glPOD_100}. Therefore, the ROM based on the dual-step POD obtains a quasi-steady solution with a few time steps. The amplitudes of the oscillations reaching a quasi-steady solution for the second mode coefficients show that the value of the amplitude is close to that of the full numerical results. Therefore, the ROM based on the dual-step POD represents the fluctuating components of the flow field obtained by the full numerical simulation.

\begin{figure}[hpbt]
   \centering
   \includegraphics[width=10.0cm]{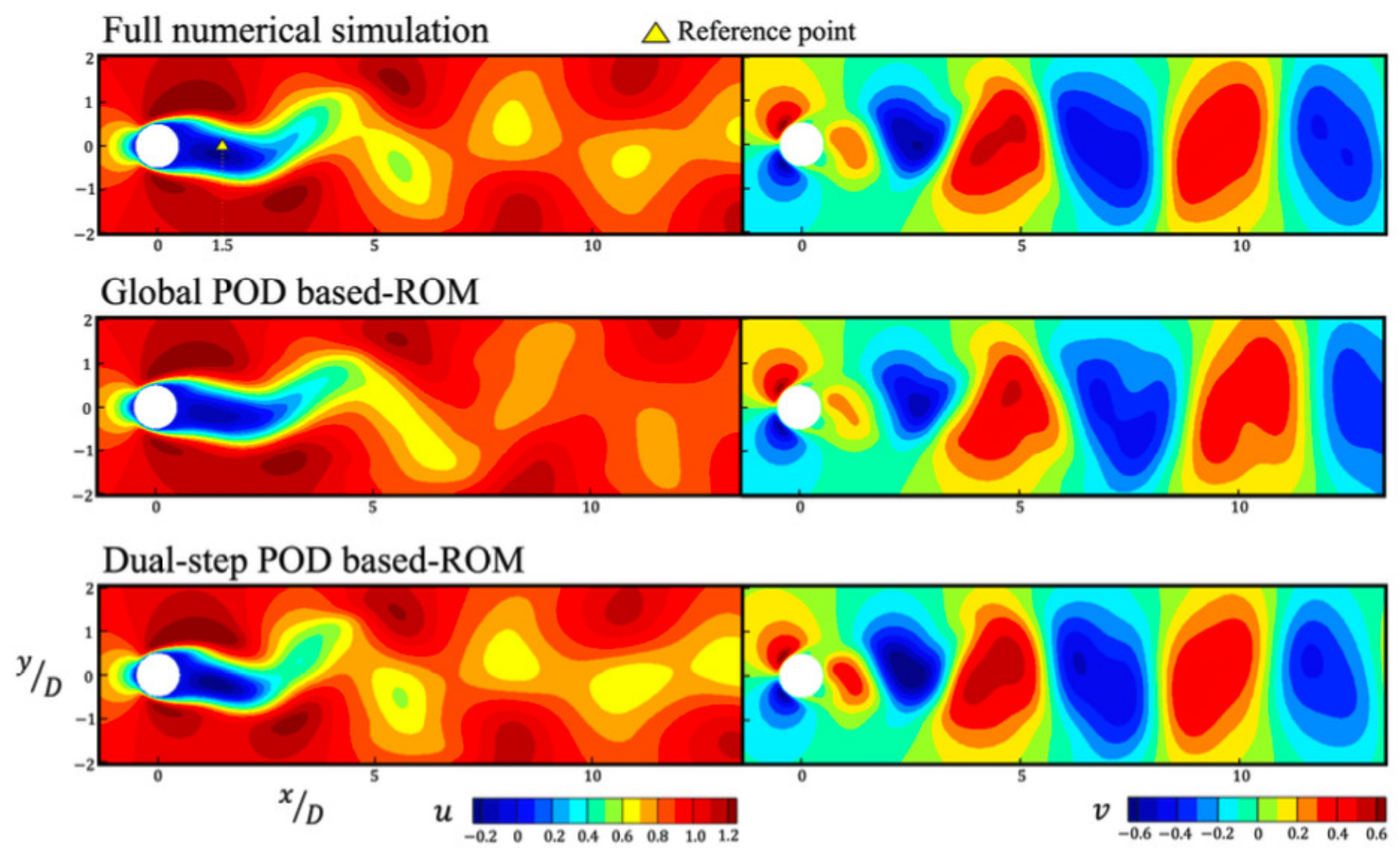}
   \caption{Instantaneous flow field around a cylinder with a Reynolds number of 100 obtained by ROM and full numerical simulation. The phase of vortex shedding is fixed at the reference point where the streamwise velocity attains its maximum.}
   \label{flowcompare}
\end{figure}

The flow field at $\text{Re}=100$ was reconstructed using the POD mode coefficients obtained from numerical simulations using the Galerkin projection approach. The instantaneous flow field at $\text{Re}=100$ reconstructed by the ROM based on the global POD and the ROM based on the dual-step POD, and the flow field obtained by the full numerical simulation are shown in Fig. \ref{flowcompare}. Each instantaneous field is the phase of maximum velocity in the streamwise direction at the reference point in the figure. 
The ROM based on the global POD results from reconstruction using $32$ POD modes and mean fields. The ROM based on the dual-step POD results from reconstruction by $10$ POD modes. These numbers of modes are determined by $\sigma_0=0.999$. The distribution of the wake flow at $x/D=0-5$ near the cylinder is in close agreement between the two ROMs and the full numerical results. The dual-step POD-based ROM results are close to the full numerical results even at $x/D>5$ in the cylinder wake. However, the global POD-based ROM results underestimate the magnitude of the velocity for $x/D>5$. Focusing on the location of the vortices behind the cylinder, the global POD-based ROM results are more backward distributed than the other two results. This is due to the POD mode used in the ROM, based on the global POD being more affected by the flow field for conditions other than $\text{Re}=100$.

\begin{figure}[!ht] 
   \centering
   \includegraphics[width=12.0cm]{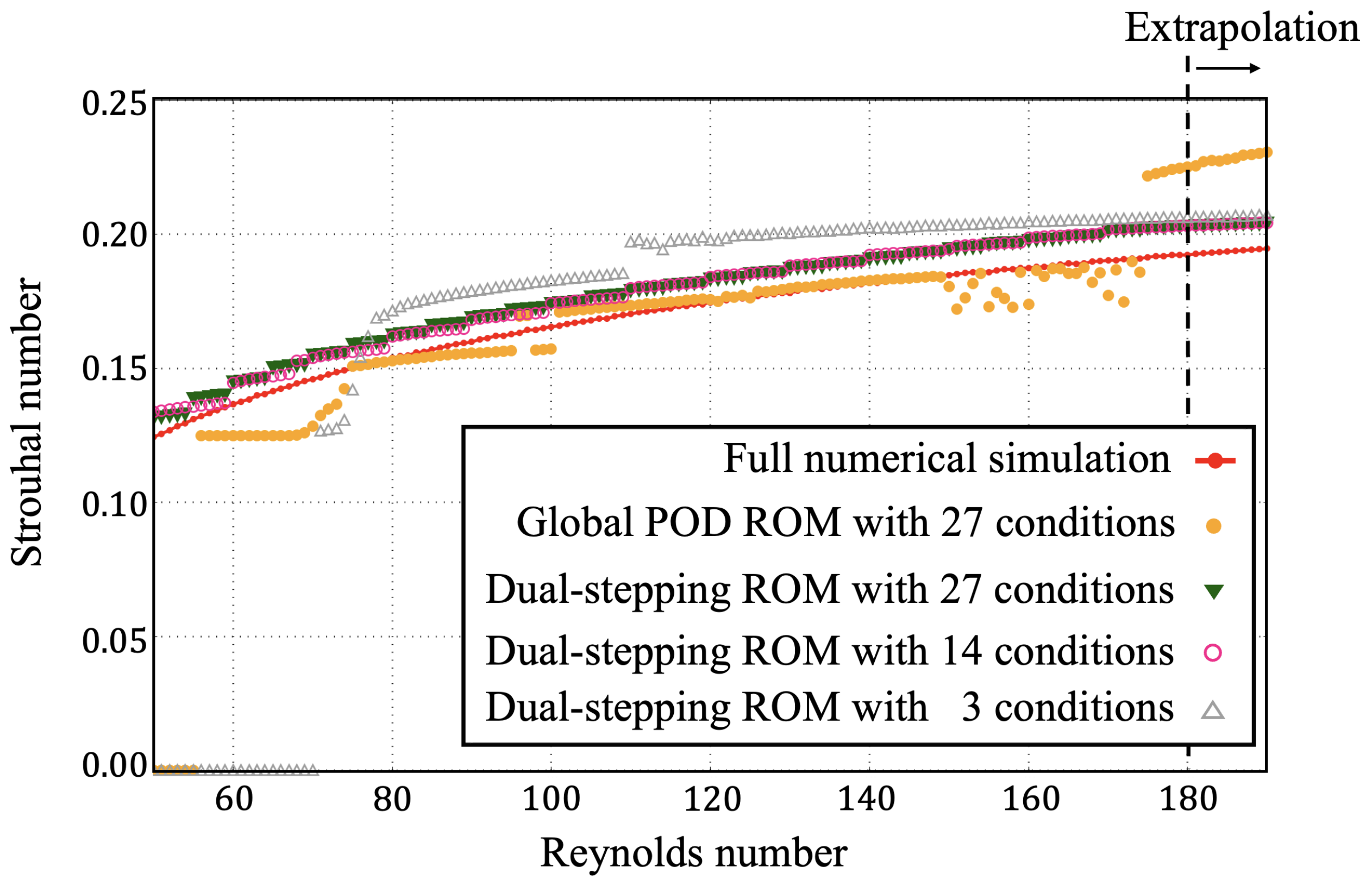}
   \caption{\textcolor{black}{Strouhal number for different Reynolds numbers obtained from full numerical simulations and ROM predictions. The ROMs are constructed using POD modes derived from datasets containing flow fields at $3$, $14$, and $27$ different Reynolds numbers.}}
   \label{coefficient_St_dualPOD}
\end{figure}

The flow field is computed by the ROM based on the dual-step POD, changing the Reynolds number of the ROM by $1$ in the range of $50$ to $180$. \textcolor{black}{Dual-step POD-based ROMs are constructed using the same datasets for global POD-based ROM.} The ROMs are performed for sufficient time to reach a quasi-steady solution for each Reynolds number, and the Strouhal number of the flow field is computed. The relationship between the Reynolds and Strouhal numbers is shown in Fig. \ref{coefficient_St_dualPOD}. \textcolor{black}{Here, the results for Reynolds numbers larger than $180$ correspond to extrapolative predictions obtained from an ROM constructed using data in the range $175$–$180$.  Although the present study is based on two-dimensional simulations, the Reynolds number of $190$ is physically close to the critical value at which the flow transitions from two-dimensional to three-dimensional behaviour\cite{Henderson, cylinder4}.} The results of the dual-step POD-based ROM are slightly larger than the Strouhal number of the full numerical simulation, but they capture the characteristics of the Reynolds-Strouhal number curve. 
\textcolor{black}{One possible reason for predicting the slightly higher Strouhal number is that the spatial distribution of the POD modes obtained through the dual-step POD represents an intermediate structure between the conditions selected in the second step and therefore does not exactly coincide with the mode corresponding to either Reynolds number. In addition, general sources of error inherent in Galerkin projection, such as modal truncation\cite{GP_deane}, the effect of the pressure term\cite{pressure2}, and the limited ability of POD modes to optimally represent nonlinear interactions, may also contribute to the observed discrepancy.}

\textcolor{black}{For Reynolds numbers larger than $180$, the extrapolative predictions exhibit similar trends to those observed within the training range, indicating that the dual-step POD-based ROM does not lose robustness under extrapolation. In contrast, the global POD-based ROM predicts significantly larger Strouhal numbers in the extrapolated regime compared with the dual-step POD-based ROM. These results suggest that the dual-step POD-based ROM provides more robust predictions over a wider range of flow conditions than the global POD-based ROM.}
Furthermore, the dual-step POD-based ROMs \textcolor{black}{using $14$ and $27$ conditions} show a quasi-steady flow field at all Reynolds numbers, in contrast to the global POD-based ROM. This contrasts with the global POD-based ROM, which resulted in a steady field when the Reynolds number is smaller than $56$.
\textcolor{black}{However, in the dual-step POD-based ROM constructed using data from only three conditions, the flow remains steady for Reynolds numbers below $70$, resulting in a limited amount of training data. Consequently, when the number of training conditions is small, the accuracy in predicting the critical Reynolds number.}
 These results show that the dual-step POD-based ROM captures the dependence of the flow field on the Reynolds number better than the global POD-based ROM \textcolor{black}{when the dataset includes a sufficiently large number of parameter conditions}. 

The CPU time required to compute the flow field is compared between an ROM based on the dual-step POD and an ROM based on the global POD. Table \ref{table4_1} shows the average CPU time required to compute 50,000 timesteps for each ROM. The CPU time for global POD-based ROM is the time required to compute 50,000 time steps in the Galerkin projection approach, starting from a given value of the Reynolds number to be reconstructed. The CPU time for the dual-step POD-based ROM is the time required to perform a second-step POD and calculate 50,000 time steps using the Galerkin projection approach. The number of time steps is determined based on the time it takes for the solution to converge in the ROM using the global POD. Note that the dual-step POD converged at about a quarter of 50,000 time steps. For each ROM, 50,000 time steps are computed for $131$ conditions with Reynolds numbers from $50$ to $180$, and the CPU time taken for the computation is recorded. The ROMs based on the dual-step PODs are faster than those based on the global PODs. Furthermore, the computation time of the ROM based on the global POD is expected to increase when more time series data are included in the dataset. Conversely, the ROM based on dual-step POD could maintain a constant computation time even as the number of time-series data in the dataset increases, since the number of modes used for Galerkin projection remains almost the same.

As shown in Fig\textcolor{black}{s}. \ref{coefficient_glPOD_100} and \ref{coefficient_long_dualPOD}, the ROM based on the dual-step POD converges to a quasi-steady solution in a smaller number of time steps than the ROM based on the global POD. Hence, an ROM based on the dual-step POD obtains a quasi-steady solution, taking less time than the table.

\begin{table}[htbp]
  \caption{Comparison of CPU time between ROM based on global POD and ROM based on dual-step POD. These ROMs are constructed using flow datasets for 27 different values of Reynolds number. The recorded CPU time includes the time required to compute the spatial derivative in the Galerkin projection and the time required for the second-step POD in the dual-step POD. }
  \label{table4_1}
  \centering
  \renewcommand{\arraystretch}{1} 
  \begin{tabular}{|c||c|}
   \hline
    ROM & CPU time\\
    \hline \hline
     Global POD & $206$\\
    \hline
    Dual-step POD & $113$\\
    \hline
  \end{tabular}
\end{table}

\section{Conclusion}\label{conclusion}
This paper presents a novel ROM \textcolor{black}{for two-dimensional cylinder flow} that improves the robustness of the conditions contained in a dataset and the computation speed for ROMs constructed by the global POD modes. The proposed ROM is constructed using a POD method that performs the global POD in two steps. The first POD is performed on the time series data for each condition in the dataset. Several flow conditions that are close to the conditions to be reconstructed are selected from conditions contained in the datasets. A second POD is performed on the two or more sets of optimal POD modes obtained from the first POD for the selected conditions. With the use of this POD, the flow field is reconstructed using the POD modes that are close to the characteristics of the flow conditions to be reconstructed. Thus, the proposed ROM effectively represents the flow field using fewer POD modes than the number of POD modes obtained from all flow fields in the dataset.

The conventional ROM using the global POD requires the inclusion of a large number of flow fields with different conditions in the dataset for the global POD to improve the robustness of the ROM. For flow around a cylinder with $\text{Re}= 50-180$, the bifurcation point from steady flow to quasi-steady flow shows different results from the full numerical results and the conventional global POD-based ROM results when the flow field with different Reynolds numbers increases in the dataset. Conversely, the proposed ROM has no bifurcation point at $\text{Re}= 50-180$. The proposed ROM represents the Reynolds-Strouhal curve of cylinder wake flow obtained by the full numerical simulation.

The number of flow fields contained in the dataset for the conventional global POD becomes large, resulting in an increase in the computation cost of computing a flow field. On the other hand, the proposed ROM reconstructs the flow with far less computation cost than the conventional global POD-based ROM when the dataset for the proposed ROM contains many conditions of flow fields. Therefore, the proposed ROM shows higher robustness and computational speed performance than the conventional ROM based on the global POD \textcolor{black}{in the case of two-dimensional cylinder flow. Although the proposed ROM framework was developed for two-dimensional flow around a circular cylinder, the methodology itself is not restricted to this specific flow configuration. Therefore, it is expected that the framework can be extended in future studies to other flow synarios, such as hydrofoils, airfoils, backward-facing step flows, and cavity flows.}


\begin{acknowledgment}
The numerical simulation for the two-dimensional Navier--Stokes equation is performed on the Supercomputer system ``AFI-NITY" at the Advanced Fluid Information Research Center, Institute of Fluid Science, Tohoku University. 

This study was partially supported by a Sasakawa Scientific Research Grant from the Japan Science Society. 
This work was partially supported by JST SPRING, Grant Number JPMJSP2114, Japan.
\end{acknowledgment}

%
% ...

\bibliographystyle{asmems4}


\appendix       
\section{\label{apenB}Time Integration Method for Galerkin Projection Approach}
The governing equation obtained by the Galerkin projection is time-integrated using the second-order Adams-Bashforth method for the first and third terms and the Crank--Nicolson method for the second term. Implicit methods are used because numerical dissipation and dispersion are unavoidable in explicit time integration, even for higher-order methods\cite{RK_dis1,RK_dis2,RK_dis3,RK_dis4}. Equation (\ref{splitlinear}) is discretized as follows
\begin{equation}
\begin{split}
&a_k^{n+1}-\frac{\Delta{t}}{2}\sum_{i=1}^{r}{a_i^{n+1}}G_{ik}
=a_k^{n}+\frac{\Delta{t}}{2}\sum_{i=1}^{r}{a_i^{n}}G_{ik}\\
&+\Delta{t}\left\{-\sum_{i=1}^{r}\sum_{j=1}^{r}F_{ijk}\frac{(3{a_i^{n}}{a_j^{n}}-{a_i^{n-1}}{a_j^{n-1}})}{2}+H_{k}\right\},
   \label{splitlinear_dis}
\end{split}
\end{equation}
where the superscript $n$ in $a^n$ represents the $n$th step coefficient. The matrix format of this equation can be rewritten by
\begin{equation}
\begin{split}
M\boldsymbol{a}=\boldsymbol{b},
   \label{matrix}
\end{split}
\end{equation}
where $M$ represents $r \times r$ matrix presented below
\begin{equation}
\renewcommand{\arraystretch}{2}
M = 
\begin{bmatrix}
1-\frac{\Delta{t}}{2}G_{11} & -\frac{\Delta{t}}{2}G_{21} & \cdots & -\frac{\Delta{t}}{2}G_{r1}\\
-\frac{\Delta{t}}{2}G_{12} & 1-\frac{\Delta{t}}{2}G_{22} & \cdots & -\frac{\Delta{t}}{2}G_{r2}\\
\vdots & & \ddots & \vdots \\
-\frac{\Delta{t}}{2}G_{1r}  &  & \cdots & 1-\frac{\Delta{t}}{2}G_{rr} \\
\end{bmatrix},
\label{compactseq9}
\renewcommand{\arraystretch}{1}
\end{equation}
$\boldsymbol{a}$ and $\boldsymbol{b}$ are $r$-dimension vector presented below
\begin{equation}
\renewcommand{\arraystretch}{2}
\boldsymbol{a} = 
\begin{bmatrix}
a_1^{n+1}\\
a_2^{n+1}\\
\vdots\\
a_r^{n+1}
\end{bmatrix},
\boldsymbol{b} = 
\begin{bmatrix}
b_1\\
b_2\\
\vdots\\
b_r
\end{bmatrix}.
\label{compactseq9}
\renewcommand{\arraystretch}{1}
\end{equation}
Here, component $b_k$ is the right-hand side of Eq. (\ref{splitlinear_dis}). In this study, Eq. (\ref{matrix}) is solved by the BiCGSTAB method since the matrix $M$ is diagonally dominated\cite{GP_deane}.

\end{document}